\documentclass[a4paper,11pt]{article}
\pdfoutput=1

\usepackage[normalem]{ulem}
\usepackage{jcappub} 
\usepackage{amsmath}
\usepackage{amssymb}
\usepackage{color}
\usepackage{graphicx}
\usepackage{comment}
\usepackage{multirow}
\usepackage[dvipsnames]{xcolor}

\usepackage[utf8]{inputenc}

\usepackage{caption}

\graphicspath{{figures/}}

\newcommand{\Om}{\Omega}

\newcommand{\be}{\begin{equation}}
\newcommand{\ee}{\end{equation}}
\newcommand{\bea}{\begin{eqnarray}}
\newcommand{\eea}{\end{eqnarray}}
\newcommand{\bean}{\begin{eqnarray*}}
\newcommand{\eean}{\end{eqnarray*}}

\newcommand{\OmM}{\Omega_{M}}
\newcommand{\OmL}{\Omega_\Lambda}

\def\msun{\,{\rm M_\odot}}

\begin{document}

\title{\boldmath Science with the space-based interferometer eLISA.\\III: Probing the expansion of the Universe using gravitational wave standard sirens} 

\author[a]{Nicola Tamanini}
\affiliation[a]{Institut de Physique Th\'eorique, CEA-Saclay, CNRS UMR 3681, Universit\'e Paris-Saclay, F-91191 Gif-sur-Yvette, France}
\emailAdd{nicola.tamanini@cea.fr}

\author[a]{Chiara Caprini}
\emailAdd{chiara.caprini@cea.fr}

\author[b,c]{Enrico Barausse}
\affiliation[b]{Sorbonne Universit\'es, UPMC Univesit\'e Paris 6, UMR 7095, Institut d'Astrophysique de Paris, 98 bis Bd Arago, 75014 Paris, France}
\affiliation[c]{CNRS, UMR 7095, Institut d'Astrophysique de Paris, 98 bis Bd Arago, 75014 Paris, France}
\emailAdd{barausse@iap.fr}

\author[d]{Alberto Sesana}
\affiliation[d]{School of Physics and Astronomy, The University of Birmingham, 
Edgbaston, Birmingham B15 2TT, UK}
\emailAdd{asesana@star.sr.bham.ac.uk}

\author[e]{Antoine Klein}
\affiliation[e]{Department of Physics and Astronomy, The University of Mississippi, University, MS 38677, USA}
\emailAdd{aklein@physics.montana.edu}

\author[f]{Antoine Petiteau}
\affiliation[f]{APC, Universit\'e Paris Diderot, Observatoire de Paris, Sorbonne Paris Cit\'e, 10 rue Alice Domon et L\'eonie Duquet, 75205 Paris Cedex 13, France}
\emailAdd{antoine.petiteau@apc.univ-paris7.fr}

\date{\today}

\abstract{
We investigate the capability of various configurations of the space interferometer eLISA to probe the late-time background expansion of the universe using gravitational wave standard sirens.
We simulate catalogues of standard sirens composed by massive black hole binaries whose gravitational radiation is detectable by eLISA, and which are likely to produce an electromagnetic counterpart observable by future surveys.
The main issue for the identification of a counterpart resides in the capability of obtaining an accurate enough sky localisation with eLISA.
This seriously challenges the capability of four-link (2 arm) configurations to successfully constrain the cosmological parameters.
Conversely, six-link (3 arm) configurations have the potential to provide a test of the expansion of the universe up to $z\sim 8$ which is complementary to other cosmological probes based on electromagnetic observations only.
In particular, in the most favourable scenarios, they can provide a significant constraint on $H_0$ at the level of 0.5\%.
Furthermore, $(\Omega_M, \Omega_\Lambda)$ can be constrained to a level competitive with present SNIa results.
On the other hand, the lack of massive black hole binary standard sirens at low redshift allows to constrain dark energy only at the level of few percent.
}

\maketitle

\section{Introduction}

Gravitational waves (GWs) are potentially a very powerful probe of the universe: they can bring information from a wide variety of phenomena, ranging from stellar processes to the collision of galaxies and the formation of structures, up to the very early stages of the evolution of the universe. Several frequency windows of gravitational radiation are now being targeted by different detectors. Present and near-future ground-based interferometers target the frequency window $10-1000$ Hz: the network of second-generation detectors includes the two advanced LIGO interferometers in the United States \cite{ligo} and the advanced VIRGO interferometer in Italy \cite{virgo}, and will be joined in the future by KAGRA \cite{kagra} in Japan and LIGO India \cite{indigo}. With the third-generation ground-based interferometer Einstein Telescope \cite{ET}, it will be possible to reach frequencies of the order of $1$ Hz, since going underground helps to partially overcome the seismic noise. On the other hand, timing of ultra-stable millisecond pulsars and mapping of the Cosmic Microwave Background (CMB) can probe much lower frequencies. Three Pulsar Timing Arrays are currently in operation -- the EPTA in Europe \cite{epta},  NANOGrav in the US \cite{nanograv} and PPTA in Australia \cite{ppta} -- sharing data under the aegis of the International Pulsar Timing Array (IPTA) \cite{2010CQGra..27h4013H}: they monitor the frequency window $10^{-8}$ to $10^{-9}$ Hz and have already placed interesting bounds on the stochastic GW background produced by  massive black-hole binaries (MBHBs)  \cite{2010CQGra..27h4013H,Shannon:2015ect}. The B-mode polarisation of the CMB gives access to the very low frequency band $10^{-18}-10^{-16}$ Hz \cite{Ade:2015tva}, targeting the GW background sourced during inflation, or similar scenarios where GWs are generated (non-causally) outside the horizon at recombination. 

The advantage of a space-based GW interferometer resides in its capability to 
reach high sensitivity in the intermediate frequency band below  $1$ Hz. This 
frequency band is unexplored so far, and very rich with both astrophysical and 
cosmological sources (for a summary, see \cite{Seoane:2013qna,Binetruy:2012ze}). 
In particular, the main target is the GW signal from MBHBs with high signal to 
noise ratio (SNR) and up to high redshift. The Laser Interferometer Space 
Antenna (LISA) program has been pursuing the goal of detecting GWs in the 
$10^{-4}-0.1$ Hz band for many years as a joint venture between the European 
Space Agency (ESA) and the United States' National Aeronautics and Space 
Administration (NASA). In 2013, ESA has approved a GW observer in space as the 
L3 mission for launch in 2034, for which the ``evolved LISA'' (eLISA) 
space-based interferometer is the main candidate \cite{elisaweb}. For this 
reason, a design study is currently in progress within ESA to define the best 
eLISA configuration from the point of view of science return and costs. 

Some characteristics of the eLISA configuration that are still 
undecided are the low-frequency noise level, soon to be tested by LISA 
Pathfinder \cite{pathfinderweb}, the number of laser links (four or six), the 
length of the interferometer arm (between one and five million km), and the 
duration of the mission (between two and five years). In the context of the 
design study, the scientific potential of each possible eLISA configuration 
obtained by varying these characteristics is under analysis. The present work is 
part of a series of papers that analyse the impact of these four key design 
choices on the scientific performance of eLISA. {\it Here, we address 
specifically the scientific goal of probing the expansion of the universe 
using MBHB GW sources as standard sirens}. The first paper of this series 
\cite{PaperI} has dealt with the cosmological evolution of MBHBs. We will refer 
to this as Paper~I in the following. The second paper \cite{Caprini:2015zlo} has 
analysed the GW stochastic background from phase transitions occurring in the 
early universe.  

The GWs emitted by a compact binary directly encode the luminosity distance of 
the system. An additional measurement of the redshift would then permit probing 
the distance-redshift relation. Binary systems emitting GWs can therefore 
constitute a very valuable tool for cosmology. Indeed, they can be thought of as 
``standard sirens'', i.e.~the gravitational analogue of standard candles such as 
type Ia supernovae (SNIa), as first pointed out in \cite{schutz} (see 
also e.g.~\cite{Holz:2005df,Cutler:2009qv}). One advantage of standard sirens 
over SNIa is that they allow for a direct measurement of the luminosity distance 
up to large redshift, unlike optical measurements, which require 
cross-calibrations of successive distance indicators at different scales. For 
this reason, they can be used in conjunction with optical observations to 
directly measure the Hubble constant, which 
cannot be obtained by SNIa observations, which are calibrated in the local Hubble flow \cite{Weinberg:2012es}. Furthermore, the luminosity of a GW source is mostly determined by 
gravitational physics (in fact, for binaries of black holes the GW luminosity depends \textit{only} on gravitational physics). As a result, GW sources present 
essentially no luminosity scatter, unlike SNIa.
Still, MBHBs have the big disadvantage that they cannot probe the distance-redshift relation by themselves.
Indeed, exactly because MBHB waveforms only depend on gravitational physics -- i.e.~on General Relativity, which has no built-in length-scale -- the two masses $m_{1,2}$ and the redshift $z$ enter
the waveforms only through the redshifted masses $m_{z 1,2}=(1+z) m_{1,2}$. Therefore, the redshift needs to be measured independently, e.g.~by optically identifying the host galaxy. This limits the capability of standard sirens to probe cosmology, as we will see in the following.  

The aim of the present analysis is to forecast the potential of several proposed eLISA configurations to measure the distance-redshift relation, 
in combination with independent redshift measurements by future optical telescopes (here, we will consider the Large Synoptic Survey Telescope (LSST) \cite{LSSTweb} and the Extremely Large Telescope (ELT) \cite{EELTweb}),
aided for the source localisation by future radio telescopes such as the Square Kilometre Array (SKA)~\cite{SKAweb}. We analyse twelve configurations, listed in the first column of Table~\ref{tab:ss} at the end of the paper: six with four laser links (L4), and six with six laser links (L6). For each configuration, the length of the arm is chosen to be one, two or five million km (A1, A2, A5), and we set the low-frequency noise level to either the LISA Pathfinder expected one (N2), or the LISA Pathfinder required one, which is ten times worse (N1). These possibilities correspond to those under investigation by ESA, except the mission duration, which we have decided to fix to five years (M5). 

First of all, one needs to model realistically the expected sources for each of the above mentioned configurations. As a starting point, following Paper~I, we use the results of semi-analytical simulations of the evolution of the BH masses and spins during the hierarchical galaxy formation and evolution~\cite{mymodel,spin_model,letter,newpaper}. This allows one to predict the rate and redshift distribution of MBHB merger events. We produce several variants of our semi-analytical model by
considering competing scenarios for the initial conditions for the massive BH population at high redshift --
 namely, a ``light-seed'' scenario in which the first massive BHs form from the 
remnants of population III (popIII) stars~\cite{MadauRees2001,volonteri_haardt_madau},
and a ``heavy-seed'' one where massive BHs form from the collapse of
protogalactic disks~\cite{2004MNRAS.354..292K,2006MNRAS.370..289B,
  2006MNRAS.371.1813L,2008MNRAS.383.1079V} --
and for the delays with which massive BHs merge after their host galaxies 
coalesce~\cite{newpaper}. 
For each variant of our model, our simulations 
allow us to produce synthetic catalogues of MBHB merger events, including all 
information about the MBHBs (masses, spins, redshift, etc.) and their host 
galaxies (mass in gas, mass in stars, etc.). In order to have enough statistics, especially at low redshift,
we produce 118 catalogues for each BH formation model and each eLISA configuration (see sections~\ref{sec:BHbinaries} and \ref{sec:statistic} for more details).

The parameters of the  MBHB systems of each catalogue are then inserted as input 
into a code that simulates the GW signal induced in the detector by the 
binaries' inspirals (cf.~section \ref{sec:detection}). For every one of the 
twelve eLISA configurations that we consider, 
the code computes the SNR of each merger event and the Fisher matrix of the 
corresponding waveform parameters. In particular, since we are to use the 
sources as standard sirens, we are most interested in the $1\sigma$ error on the 
luminosity distance $\Delta d_L$, and on the sky location $\Delta \Omega$. Among 
the MBHB merger events, we select those that have ${\rm SNR}>8$ and $\Delta 
\Omega <10\,{\rm deg}^2$. This guarantees that the events are measured with a 
sky location error sufficiently small to allow one to detect a counterpart, if 
it is present. We choose 10 ${\rm deg}^2$ because it corresponds to the field of 
view of the LSST survey \cite{LSSTweb}. 

The following step is to select, among these events, those that are likely to provide a detectable optical counterpart (for the redshift determination). We consider a two-case scenario. On the one hand, the optical emission from the counterpart may be bright enough to be directly detectable by an optical survey such as LSST. In this case, we have a direct determination of the GW source redshift. On the other hand, the counterpart may take the form of a radio jet or flare. In this case, the counterpart may first be detected in the radio band by the SKA \cite{SKAweb}. When the counterpart has been localised in the radio, the host galaxy can be identified and its redshift estimated either spectroscopically or photometrically (depending on the galaxy's luminosity) using an optical telescope such as the ELT \cite{EELTweb} (or possibly the James Webb Space Telescope, JWST \cite{JWSTweb}). 

Several authors (see, e.g.,~\cite{2002ApJ...567L...9A,2010Sci...329..927P,Giacomazzo:2012iv}) suggested that optical/radio emission is likely to happen at merger in a transient fashion (see however \cite{2014PhRvD..90j4030G,2015MNRAS.447L..80F} for models in which the electromagnetic emission can take place also continuously during the inspiral phase). In order to be able to detect the associated electromagnetic emission, which is essential to use the MBHB as standard siren, ideally optical/radio telescopes should be pointed in the direction of the event prior to merger. The most conservative procedure we can conceive is therefore to select the events with ${\rm SNR}>8$ and $\Delta \Omega <10\,{\rm deg}^2$ on the basis of the eLISA detection and parameter estimation {\it during the inspiral phase only}. It is reasonable to expect that, by the time eLISA flies, real-time data analysis should be feasible, allowing one to pin down the sky location of the MBHB as it spirals in. Among the observed MBHBs, there will be some for which, just prior to coalescence, the sky location accuracy is better than $10\, {\rm deg}^2$. As soon as the event has been localised in the sky with this precision, telescopes are alerted and point in that direction, looking for a distinctive flare occuring at merger. This procedure is the one that maximises the chances to detect the electromagnetic emission from the counterpart. However, not many events are expected to be seen with ${\rm SNR}>8$ and at the same time to be localised with $\Delta \Omega <10\,{\rm deg}^2$ on the basis of the analysis of the inspiral phase only. Moreover, most of the localisation accuracy builds up in the last few days of the inspiral, as first pointed out in \cite{Kocsis:2007hq} and subsequently confirmed by a more accurate analysis in \cite{Lang:2007ge}. The MBHB must be followed until close to coalescence before the necessary information on the sky location of the event can be communicated to the telescopes.

It is reasonable to conceive also another strategy, whereby the number of MBHBs that can be observed with ${\rm SNR}>8$ and $\Delta \Omega <10\,{\rm deg}^2$ would increase significantly: this consists in {\it adding the merger and ringdown portions of the waveforms} in the determination of the SNR and in the estimate of the sky location to be provided to the telescopes. By doing so, new events would be included, which would be otherwise left out when performing the ${\rm SNR}>8$ and $\Delta \Omega <10\,{\rm deg}^2$ cuts considering the inspiral phase only. This happens in particular for the most massive binaries, or for those merging at higher redshift. Clearly, in this procedure the telescopes can be pointed only {\it after the merger} to look for distinctive signatures. Therefore, whether one can actually take advantage of the merger and ringdown in the sky localisation depends on the nature and the timing of the counterpart (remember we assume real-time data analysis for eLISA). If there is a delay between the merger and the flare, or if the electromagnetic signal is persistent and peculiar enough that it can be confidently identified also minutes to hours after merger, then the above described procedure is feasible.


We therefore consider both a  ``conservative scenario'', in which the selection of the events with ${\rm SNR}>8$ and $\Delta \Omega <10\,{\rm deg}^2$ is performed using only the inspiral phase, and an ``optimistic scenario'', in which we include the merger and ringdown phases for the selection of these events.
In the latter case, we also include a de-lensing of a factor of two, since by the time eLISA flies one might be able to rely on quite accurate lensing maps (for details, see section \ref{sec:statistic}). As we will see, in this optimistic scenario the number of useful standard sirens substantially increases (especially for the heavy-seed models), thereby greatly improving the precision with which the cosmological parameters can be measured. 

Note that once the optical/radio counterpart has been identified at merger, the exact sky location of the binary is known. As a consequence, the error on the luminosity distance $\Delta d_L$ is significantly reduced by repeating the Fisher matrix analysis after the merger, when two parameters (the sky location angles) can be fixed and the merger and ringdown parts of the waveform can be added to the Fisher matrix analysis. This procedure improves the measurement of the cosmological parameters.

In summary, our catalogues of sources that can be used to investigate the distance-redshift relation are composed by MBHBs that can be detected by a given eLISA configuration 
with ${\rm SNR}>8$ and $\Delta \Omega <10\,{\rm deg}^2$, either with or without merger and ringdown, and whose counterpart is furthermore visible either directly in the optical by 
LSST, or in the radio by SKA (with follow-up optical observations of the host galaxy by ELT). 

\begin{figure}[tbp]
\centering
\includegraphics[width=.65\textwidth]{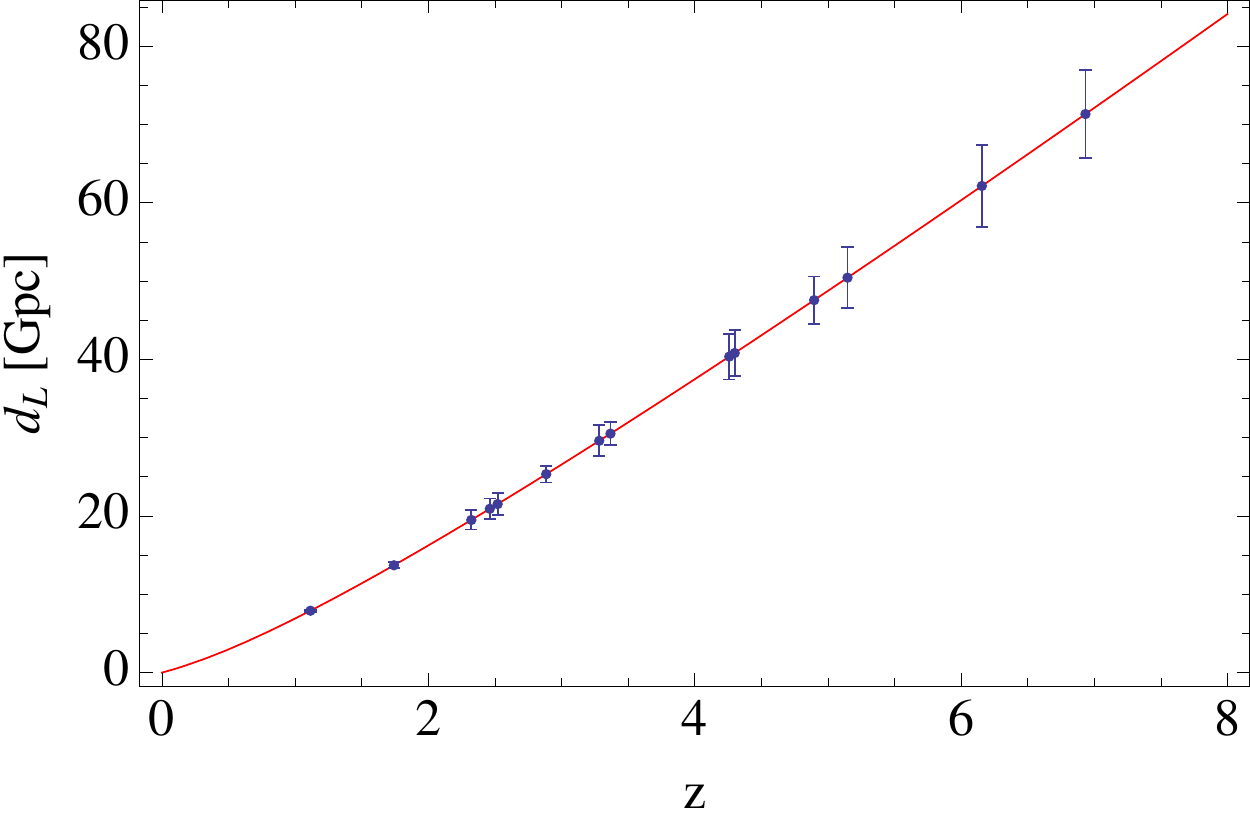}
\caption{Example of data points in the distance-redshift diagram coming from MBHB standard sirens simultaneously detectable with eLISA and an electromagnetic telescope as described in the introduction (here we choose one random catalogue for N2A5M5L6 in the model ``heavy seeds no delay''). Note that the eLISA mission will be able to map the expansion of the universe in the range $z \gtrsim 2$, which cannot be tested by SNIa or other current probes. On the other hand, few standard sirens will be available at lower redshift.}
\label{fig:data}
\end{figure}

Each MBHB is a point in the plane $(z, d_L(z))$, with a 1$\sigma$ error on the luminosity distance $\Delta d_L$ comprising of different components. First, the ``experimental'' error of the eLISA measurement, either with or without merger and ringdown, further reduced by the exact localisation of the source allowed by the counterpart detection. Second, we add to this ``experimental" error those due to weak-lensing and peculiar velocities, $\sigma_{\rm lens}(z)$ and $\sigma_v(z)$ \cite{Holz:2005df,camille}. As mentioned, in the optimistic scenario (which accounts for merger and ringdown), we further reduce the weak-lensing error by a factor of two. This is to mimic the effect of a certain amount of de-lensing, which we assume will be feasible in 2034 (see discussion in section \ref{sec:statistic}). Third, we add the error on the redshift determination as follows: if the redshift can be determined spectroscopically, we assume that the error is negligible for the purpose of our analysis; however, if the host galaxy is faint and the measurement has to be done photometrically, we propagate a non negligible redshift error on the luminosity distance error, as explained in section \ref{sec:statistic}. An example of simulated data points with their error bars is given in Fig.~\ref{fig:data}. 

Equipped with the catalogues of standard sirens, in order to assess the capability of eLISA to probe cosmology we use a Fisher matrix technique. We assume that the luminosity distance $d_L(z)$ depends on five cosmological parameters: the matter density parameter $\Omega_M$ at $z=0$, the cosmological constant/dark energy density parameter $\Omega_\Lambda$ at $z=0$, the Hubble constant $H_0$ at $z=0$, the equation of state of dark energy $w_0$ at $z=0$, and its evolution rate $w_a$ (for details, see section \ref{sec:cosmology}). For each catalogue  of simulated eLISA observations, we calculate the five-dimensional Fisher matrix for these parameters. 
For each Fisher matrix, we calculate a Figure of Merit (FoM), as defined in section \ref{sec:statistic}. We therefore have a distribution of FoM, from which we can extract a representative FoM, for each MBHB formation model and each eLISA configuration.  

We analyse five cosmological models. The first one is the full five-parameter 
model, and we find that no eLISA configuration is sensitive enough to provide 
meaningful constraints on it (note that also other current cosmological probes, if not combined together, are unable to give interesting constraints on a five-parameter cosmological model). This is mainly because of two 
reasons: first, all the simulated catalogues are poor in low-redshift events, 
where the effect of the cosmological constant/dark energy is the largest. 
Second, there are strong degeneracies among these parameters in the way they 
enter the distance-redshift relation, in particular among the triplet 
$(\OmM,\,\OmL,\,H_0)$ and the couple $(w_0,\,w_a)$. Therefore, in the same 
spirit as in previous analyses utilising standard sirens \cite{Zhao:2010sz}, 
SNIa \cite{Betoule:2014frx} or BAO \cite{Aubourg:2014yra}, we consider other 
cosmological models where some of the parameters are fixed. For instance, we fix 
$w_0=-1\,,~w_a=0$ (pure cosmological constant) and assess the capability of the 
different eLISA configurations to test the presence of a cosmological constant 
in a universe with curvature (i.e.~with $\OmM+\OmL\neq1$). Furthermore, we also 
fix the curvature to zero ($\OmM+\OmL=1$) and test the standard $\Lambda$CDM 
scenario. Also, to test the presence of a dynamical dark energy (DDE) component, 
we fix $(\OmM,\,\OmL,\,H_0)$ and study the ability of several eLISA 
configurations to constrain $(w_0,\,w_a)$. All the results are presented in 
section \ref{sec:results} for the two scenarios: conservative, without 
accounting for merger and ringdown, and optimistic, accounting for merger and 
ringdown plus a de-lensing of a factor of two. 

The rest of the paper is organised as follows: in section \ref{sec2} we present 
the different detector configurations; in section \ref{sec:BHbinaries},
we present the model that we use to simulate the MBHB population and its cosmological evolution; 
in section \ref{sec:detection}, we explain how we analyse the detectability of the GW emission from MBHBs and estimate their parameters; in section \ref{sec:counterparts}, we present our assumptions for modelling the electromagnetic counterparts; in section \ref{sec:cosmology}, we present the cosmological theory underpinning our study; in section \ref{sec:statistic}, we review our statistical methods; in section \ref{sec:results}, 
we present our results for the projected eLISA constraints on the cosmological parameters; finally, in section \ref{sec:FoM} we compare the performance of the different eLISA configurations. We conclude in section \ref{sec:conclusions}.

\section{Detector configurations}
\label{sec2}

As mentioned in the introduction, following Paper~I we analyse 
twelve detector configurations, corresponding to considering \textit{(i)}
possible arm lengths of $1,2,5$ Gm (A1, A2, A5 respectively);
\textit{(ii)} possible low frequency noise levels corresponding either
to the LISA Pathfinder expected noise (N2) or ten times worse (N1), and \textit{(iii)} four (L4) or six (L6) links in the satellite constellation, corresponding to a single or two (virtually) independent Michelson interferometers. In more detail,
laser powers are fixed to 0.7W (2W) for the A1 (A2 and A5) configurations, while
the telescope mirror diameter is chosen to be 25, 28 and 40 cm for A1, A2, and A5 respectively.
Analytic fits for the corresponding sensitivity curves are provided in Paper~I, 
and have been tested
against the numerical sensitivity curves produced with the {\rm LISACode}
simulator. 

In addition to the instrumental noise, we also account for the astrophysical
foreground from compact white dwarf binaries in our Galaxy. These binaries are
expected to be of the order of millions and to emit almost monochromatic GW signals with
uncorrelated phases. As such, these sources produce  an
unresolved stochastic foreground at frequencies $\sim$ mHz, and
only sufficiently loud sources can be extracted singularly~\cite{Blaut:2009si,Littenberg:2011zg,Crowder:2007ft}. A fit (based on population synthesis models~\cite{Nelemans:2003ha}) to
this unresolvable stochastic foreground noise is presented in Paper~I.

\section{Catalogues of BH binaries}
\label{sec:BHbinaries}

Our models for the population of MBHB are the same as those of Paper~I, which
are in turn based on the semi-analytical galaxy formation model of \cite{mymodel} (with later incremental improvements described in \cite{spin_model,letter,newpaper}).
We refer to those papers for more details, and here we limit ourselves to
reviewing the main features of our models.

\begin{figure*}[th]
\centering
\includegraphics[width=14cm,angle=0.]{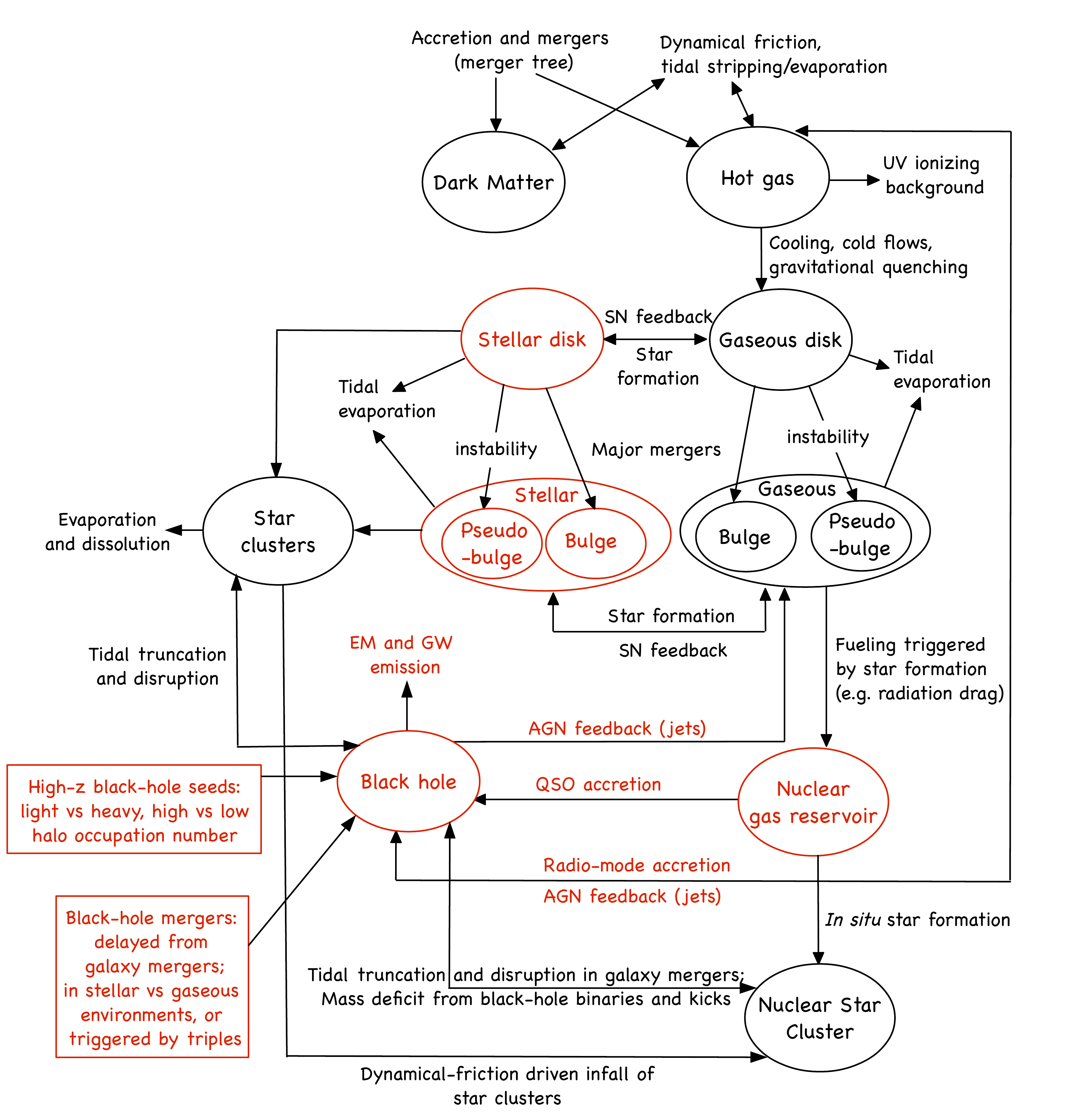}
\caption{Graphical representation of the model of~\cite{mymodel} (with the improvements described in~\cite{spin_model,letter,newpaper}). The red boxes on the left 
highlight the prescriptions (BH seeds and delays) for which multiple options are considered in this work. Also marked in red are the
elements of the model that will be used to predict the presence of electromagnetic counterparts in Sec. \ref{sec:counterparts}.
This figure has been adapted from Paper~I.
\label{fig:model}}
\end{figure*}

Our semi-analytical galaxy formation model allows us to track the cosmological evolution of the galactic baryonic structures (e.g.~the hot, primordial intergalactic medium;
the cold, chemically enriched interstellar medium -- in both its disk and bulge components; the galaxy's stellar disk and spheroid; the gas in the nuclear region; the nuclear star cluster and the
massive BH) along dark matter merger trees produced with an extended Press-Schechter formalism, modified to match the results of N-body simulations~\cite{mergertree}.
Our model accounts for a plethora of gravitational and non-gravitational interactions between these components, as represented schematically in Fig.~\ref{fig:model}.
Two aspects of our model that are key to making reliable predictions for the number of massive BH mergers observable by eLISA are
highlighted in red boxes in Fig.~\ref{fig:model}: the black hole seeding
at high redshift, and the delays between the merger of two galaxies and that of the massive BHs they host. 
Also shown in red circles are other aspects of our model that we will use in this paper to assess the presence or absence of an electromagnetic counterpart, and
the possibility of measuring the redshift.

In more detail, we assume two competing models for the high-redshift seeds from which massive BHs grow, namely a ``light-seed'' scenario, and a ``heavy-seed'' one.
In the former, the massive BHs are assumed to grow from 
the remnants of population III (popIII) stars forming at
$z\approx 15-20$~\cite{MadauRees2001,volonteri_haardt_madau} in the most massive, rarest dark matter halos. 
We assume that these remnants have mass of roughly $2/3$ of that of the progenitor popIII star. The latter is drawn from a log-normal initial  mass function centred at $300 M_\odot$ (with a r.m.s.~of 0.2 dex
and an exclusion region
between 140 and 260 $M_\odot$,
where popIII stars are subject to the electron-positron pair instability
and thus do not form a BH as they explode as supernovae)~\cite{2002ApJ...567..532H}.
In the latter, heavy-seed model, the massive BHs are instead assumed 
to form from the collapse (driven for instance by bar instabilities) of
protogalactic disks. 
This process could produce a large influx of gas to the galactic nucleus, as a result of which
BH seeds with masses $\sim 10^5 M_\odot$ might form at $z\sim 15-20$.
Several versions of this heavy-seed scenario 
have been proposed~\cite{2004MNRAS.354..292K,2006MNRAS.370..289B,
  2006MNRAS.371.1813L,2008MNRAS.383.1079V}, but here we
follow Paper~I
and adopt the model of~\cite{2008MNRAS.383.1079V}.
This model has in particular a tuneable parameter $Q_c$ (which is the critical Toomre 
parameter regulating the onset of the bar instability of the protogalactic disks), which
regulates the fraction of halos hosting a BH seed at high redshift, and whose plausible values are $2\lesssim Q_c\lesssim 3$.

As mentioned above, also crucial for predicting the number of massive BH mergers is the time (``delay'')
at which they take place after a galaxy merger. Indeed, when two galaxies coalesce, the massive BHs do not yet
form a bound binary, but rather interact separately with the stellar background via dynamical friction,
as a result of which they slowly fall toward the center of the newly formed galaxy. When their relative velocity exceeds 
the velocity dispersion of the stellar background, the two BHs form a ``bound'' (or ``hard'') binary. At this
stage of the system's evolution, dynamical friction becomes ineffective, and the binary's evolution starts being driven by
three-body interactions with the stars of the galactic nucleus.

It is currently still unclear if these three-body interactions are enough to bring the massive BHs to merger. 
Indeed the BH binary may stall at separations much larger than those ($\lesssim 10^{-3}$ pc)
at which GW emission drives the system to coalescence within a Hubble time. 
This is known as the ``final-parsec problem''~\cite{begelman}. Recent work seems however 
to suggest that a triaxial galactic potential (arising e.g.~from a recent galaxy merger)~\cite{yu,lastPc1,2014CQGra..31x4002V,lastPc2,2015arXiv150505480V} or galaxy rotation~\cite{2015arXiv150506203H}
could enhance the effect of three-body interactions, thus causing the binary to merge in a few Gyr.
The presence of gas also has a strong impact on the evolution of MBHBs. Indeed,
if sufficient gas is present in the nuclear region, planet-like migration
may drive the BH binary to coalescence within  $\sim 10^7-10^8$ yr~\cite{bence,2014SSRv..183..189C}
(although this scenario is not immune from possible complications, see e.g.~\cite{2009MNRAS.398.1392L}).
Furthermore, if a BH binary stalls for too long after the galaxy merger that originally formed it, 
a new galaxy merger may take place, which may bring a third massive BH to the system. Kozai resonances
may then trigger the merger of the two heaviest BHs (and ejection of the lightest) in $\sim 10^8$ yr~\cite{hoffman}, at least 
for BH masses $\gtrsim 10^6-10^7 M_\odot$. Below this threshold, this
process might be ineffective at causing mergers, especially if the third BH
is much lighter than the inner binary (in which case it would be ejected before it can trigger
the merger of the inner binary).

A more detailed description of the implementation of these delays in our model is beyond the scope of the present paper, and we refer 
the interested reader to~\cite{newpaper}. In order to highlight their impact on our results, in this paper we will consider both models in which these delays
are included, and one in which these delays are absent (i.e.~the massive BHs coalescences take place at the same time as galaxy mergers). More specifically, we
follow Paper~I and work with the following representative models:

\noindent {\bf 1)~Model popIII}:
A ``realistic'' light-seed model with delays included. As mentioned in Paper~I, 
the corresponding light-seed model with no delays included would 
present BH merger rates differing only by a factor $\sim 2$.

\noindent {\bf 2)~Model Q3d}: 
A ``realistic'' heavy-seed model with delays included. The critical
Toomre parameter is set to $Q_c=3$. As mentioned in Paper~I, choosing $Q_c=2$
would only decrease the BH merger rates by a a factor $\sim 2$.

\noindent {\bf 3)~Model Q3nod}: The same as model {Q3d}, but
with no delays. Paper~I showed indeed that unlike in the light-seed scenario,
neglecting the delays dramatically increases the BH merger rates in the heavy seed
scenario with $Q_c=3$. As such, this model is to be considered an ``optimistic''
(upper bound) scenario for eLISA merger rates.

We utilise the set of simulations of Paper~I, which describe, for each of these three models, the 
cosmological evolution of about 1300
galaxies/galaxy clusters, with dark matter halo masses from $10^{10} M_\odot$ to $10^{16} M_\odot$
at $z=0$.
These simulations are used to produce synthetic catalogues of eLISA 
observations, i.e.~for each of the three models above we
produce $580$ years worth of data, 
i.e.~$118$ five-year catalogues.

Finally, we note
that by tracking self-consistently the various baryonic
components of a galaxy  -- including the masses, spin magnitudes and orientations
for the massive BHs --, as well as their interactions (e.g.~feedback, accretion, etc.; cf.~Fig.~\ref{fig:model})
{\it our model allows us to predict whether a BH merger takes place
in a gas-rich or gas-poor nuclear environment, and whether it is likely to have an
electromagnetic counterpart or not}. More details about our prescriptions for these counterparts will
be presented in section~\ref{sec:counterparts} below.

\section{eLISA detection and waveform parameter estimation}
\label{sec:detection}

Our analysis of the GW signal is the same as in Paper~I, and here we only provide a brief review of our approach. 

The response of the detector is modelled as in \cite{cutler-98}, and we utilise a gravitational waveform model with generic precessing spins for the inspiral phase. This is
the ``shifted uniform asymptotics'' (SUA) waveform of \cite{kcy-14}, which also includes higher-order modes besides the dominant one.
This waveform model is used both to calculate  the signal-to-noise ratio (SNR) of each MBHB in our catalogues, 
and for the parameter estimation, i.e.~to calculate, via a Fisher matrix approach, the errors on the binary's intrinsic parameters (redshifted masses $m_{1z}$ and $m_{2z}$,
spin magnitudes $\chi_{1}$ and $\chi_2$, the four angles defining the spin orientations, and the luminosity distance $d_L$)
and on the extrinsic ones (sky position, inclination, polarization angle, time 
and GW phase at coalescence).
Whenever appropriate to account for the effect of the merger and ringdown (i.e.,
in the optimistic scenario described in the introduction; or  after a counterpart has been identified based on the inspiral alone, in the conservative scenario)
we further correct the results of this analysis by using results 
obtained with aligned (or anti-aligned) spin inspiral-merger-ringdown (IMR) ``PhenomC'' 
waveforms~\cite{santamaria-et-al-10} and a set of dedicated precessing-spin IMR 
hybrid waveforms (cf.~the Appendix of Paper~I).
This correction becomes increasingly significant for heavier MBHBs. In fact, a more massive binary emits at lower frequencies, with the net result that only a small portion of the late inspiral falls within the eLISA band. In this case the largest contribution to the total SNR comes from the merger and ringdown portion of the signal, which therefore becomes critical for a good estimate of the parameters of the system.

In more detail, the merger and ringdown correction is implemented as follows. 
The ``PhenomC'' waveforms are used to rescale the inspiral-only SNR obtained from the SUA waveforms, which results in an SNR gain $\cal R$ due to the merger and ringdown. This gain can be as high as ${\cal R}\sim10-20$ at high masses $m_{1z}+m_{2z}\sim10^7 M_\odot$. Since the ``PhenomC'' waveforms assume aligned (anti-aligned) spins, 
 we take the projection of the spins of our catalogues (which have generic orientations) along the orbital angular momentum, and identify these projections
with the spins appearing in the ``PhenomC'' waveforms. We stress that this is a good approximation at least for almost aligned/anti-aligned spins, or for small spin magnitudes.
Indeed, most of the events in our catalogues involve binaries with almost aligned spins.

In order to correct our Fisher matrix analysis for the merger and ringdown effects,  one can again utilise the SNR gain $\cal R$. In more detail, a reduced Fisher matrix analysis applied to the  dedicated precessing-spin IMR hybrid waveforms mentioned above shows that the errors in the luminosity distance ($\Delta d_L/d_L$) and the sky location ($\Delta \Omega$) scale as $\Delta d_L/d_L\propto {\cal R}^{-1}$ and  $\Delta \Omega\propto {\cal R}^{-2}$ for six-link configurations, while for four-link configurations parameter-estimation degeneracies limit the gain to $\Delta \Omega\propto {\cal R}^{-1}$ and 
to a factor $\sim 2$ improvement (roughly independent of $\cal R$) for  $\Delta d_L/d_L$.
We note that because the masses and spins are fixed in our hybrid waveforms, we cannot determine how the errors on these parameters are
affected by the inclusion of the merger and ringdown. Nevertheless, since the 
merger and ringdown phase
is characterised by the mass and spin of the final BH remnant (and \textit{not} by the masses and spin of the binary's components),
we do not expect the inclusion of merger and ringdown effects to improve 
significantly our estimation of $m_{z1,2}$ and $\chi_{1,2}$.

\section{Observation of electromagnetic counterparts}
\label{sec:counterparts}

We first consider a conservative scenario, in which we select all the events 
with ${\rm SNR} > 8$ (eLISA detections) and $\Delta\Omega < 10\,{\rm deg}^2$, 
both obtained by analysing only the inspiral part of the waveform. Therefore,
this choice selects events with potential for a coincident detection by electromagnetic probes planned to be in operation at the same time as eLISA, on the basis of the sky location obtained by the inspiral waveforms alone. It also assumes that an electromagnetic counterpart is promptly triggered at the merger, and that the eLISA data analysis can be performed in real time or with low latency compared to the counterpart trigger.
On the other hand, we also consider an optimistic scenario, in which we assume that the counterpart's trigger is either sufficiently delayed to
allow for the merger and ringdown parameter estimation be completed first, or that the counterpart still has distinctive features and/or variability 
by the time the merger and ringdown have been analysed. 
In practice, in the optimistic scenario we include the effect of the merger and ringdown 
already in the selection of the events 
with ${\rm SNR} > 8$ and $\Delta\Omega < 10\,{\rm deg}^2$.

By the time eLISA is launched, there will be electromagnetic telescopes with adequately large field of view and sensitive enough to possibly detect faint signals coming from relatively low-mass systems merging at cosmological distances. We consider here two observing strategies: the first involving the Large Synoptic Survey Telescope (LSST \cite{LSSTweb}) and the second leveraging on the Square Kilometre Array (SKA \cite{SKAweb}) in combination to the Extremely Large Telescope (ELT \cite{EELTweb}). 

\subsection{LSST}

LSST is an optical $8.4\,$m telescope, covering the $u,g,r,i,z,y$ photometric bands, currently scheduled to be operational by 2022. Its field of view of 9.6 deg$^{2}$ makes it particularly suitable to observe sky regions comparable with potential GW sky-error boxes. Little is known about optical signatures potentially associated to a MBHB merger, one of the main uncertainties being the availability of gas in the vicinity of the system in the standard circumbinary disk picture (see, e.g., \cite{2012AdAst2012E...3D} and references therein). Without invoking a specific model, we just assume
that in the presence of a sufficient amount of gas, a quasar-like luminosity flare (at some fraction of the Eddington luminosity) is powered during a merger. 

Our galaxy evolution model quantifies the gas reservoir surrounding the binary at the moment of merger, from which we can consistently compute a plausible accretion rate and a bolometric flare luminosity (details are presented in Appendix \ref{app:luminosity}). The bolometric luminosity is then converted into apparent magnitude $m$, in a given observed band, via \cite{1990hsaa.book.....Z}
\begin{equation}
	L_{\rm bol} = 3.02 \times 10^{35-\frac{2}{5} M}\, {\rm erg}/{\rm sec} \,
\end{equation}
\begin{equation}
	m = BC + M - 5 + 5 \log_{10} \left(\frac{d_L}{\rm pc}\right) \,,
\end{equation}
where $d_L$ is the luminosity distance (obtained with a fiducial cosmology), $M$ is the absolute magnitude and $BC$ is the bolometric correction (which depends on the observed band \cite{1968ApJ...154...21O}). For standard quasars, a (linear) bolometric correction of 3-to-10 applies to single filter bands ranging from 145nm to 510nm \cite{2012MNRAS.422..478R}. For a given rest-frame band, the correction obviously depends on the redshift. We notice, however, that LSST has a comparable sensitivity in different bands, which narrows the gap between observed and bolometric light. Without entering in cumbersome details which are beyond the scope of this study, we adopt a simple fiducial correction of $BC=1$.

With a $30\,$s exposure, LSST will reach a limiting (5$\sigma$) apparent magnitude of $m\sim24.5$ in the $u, g, r$ bands, roughly covering the 300nm-700nm wavelength range \cite{2008arXiv0805.2366I}. However, the merger of a fiducial $10^6\msun$ MBHB lasts for several minutes. One can therefore observe in subsequent pointings of 10 minutes around the time of the merger, to see if a flare occurs in one of the galaxies in the field of view at the right time. This roughly brings an improvement to the limiting magnitude of log$_{2.5}\sqrt{600/30}\approx1.5$.  We therefore set $m_{\rm LSST}=26$ as a fiducial detection limit.
Combining the equations above we obtain the detection condition 
\begin{equation}
  m = 82.5 + BC - \frac{5}{2} \log_{10} \left( \frac{L_{\rm bol}}{3.02} \frac{\rm sec}{\rm erg} \right) + 5 \log_{10}\left(\frac{d_L}{\rm pc}\right) \leq m_{\rm LSST}
	\label{eq:LSST_cond}
\end{equation}
The identified sources can then be followed-up with a spectrograph to get a virtually exact (for our purposes) spectroscopic redshift.

\subsection{SKA+ELT}

As a second example, we consider the observability of merging MBHBs in the radio band. Performing  force free simulations of
 merging MBHBs in an external magnetic field, \cite{2010Sci...329..927P} found dual jets being launched close to merger, with luminosity 
\begin{equation}
	L_{\rm flare} = \epsilon_{\rm edd} \epsilon_{\rm radio} (v/v_{\rm max})^2 q^2 L_{\rm edd} \,.
\end{equation}
Here, the factor $(v/v_{\rm max})^2$ describes the luminosity evolution as the MBHB inspirals ($v_{\rm max}=c/\sqrt{3}$ -- where $c$ is the speed of light -- is the circular speed at the innermost stable circular orbit for a binary of non-spinning BHs), $q=M_2/M_1\leq 1$ is the binary's mass ratio, $\epsilon_{\rm radio}$ (which we set to a fiducial value of 0.1, cf.~\cite{2011ApJ...743..136O}) is the fraction of electromagnetic radiation emitted in the radio band (i.e.~a radio-to-bolometric luminosity correction), and the Eddington ratio is defined as $\epsilon_{\rm edd} = L_{\rm bol}/L_{\rm edd}$. Although \cite{2010Sci...329..927P} found $\epsilon_{\rm edd}=0.002$, a subsequent investigation \cite{2012ApJ...749L..32M} additionally identified a brighter quadrupolar contribution to the emission implying $\epsilon_{\rm edd}=0.02$. Moreover, as noted in \cite{2011ApJ...734L..37K}, the magnetic field strength $B=6\times10^4$G$m_8^{1/2}$ ($m_8$ being the mass of the MBHB normalized to $10^8 M_\odot$) adopted in the aforementioned studies is smaller than the maximum field strength $B=10^6$G$(\alpha m_8)^{-7/20}$ created by magneto-rotational instabilities at the inner edge of a thin circumbinary disk described by the Shakura-Sunyev viscosity parameter $\alpha$ (generally found in the range $0.1<\alpha<1$ in MHD simulations \cite{2012ApJ...749..118S}). The radio luminosity of the resulting jet can therefore be correspondingly larger. Moreover, general relativistic magnetohydrodynamic simulations of magnetized  plasmas around MBHBs~\cite{Giacomazzo:2012iv} find that as a result of accretion onto the binary, the magnetic field gets amplified by $\sim 2$ orders of magnitude (hence the electromagnetic signal is enhanced by a factor $\sim 10^4$) relative to the force-free simulations of \cite{2010Sci...329..927P}.
Given these uncertainties, we simply compute $\epsilon_{\rm edd}$ from the gas accretion rate at merger (directly provided by our model, see Appendix \ref{app:luminosity}), but we impose a floor at $\epsilon_{\rm edd}=0.02$.

Besides this effect, which is due to the twisting of magnetic field lines by the rapidly inspiralling MBHB, accretion also powers standard radio jets through the Blandford-Znajeck effect~\citep{1977MNRAS.179..433B}, with luminosity dependent on the mass accretion rate. Closely following \cite{mymodel,2001ApJ...548L...9M} we use the jet luminosity 
\begin{equation}
	L_{\rm jet} = \begin{cases}
		 10^{42.7} {\rm erg}\ {\rm s}^{-1} 
\left(\frac{\alpha}{0.01} \right)^{-0.1} m_9^{0.9}
\left(\frac{\dot{m}}{0.1}\right)^{6/5}(1 + 1.1 a_1 + 0.29 a_1^2),  &\text{if } 10^{-2} \leq \epsilon_{\rm edd} \leq 0.3  \,, \\
		 10^{45.1} {\rm erg}\ {\rm s}^{-1}
\left(\frac{\alpha}{0.3}\right)^{-1} m_9 
\left(\frac{\dot{m}}{0.1}\right) g^2(0.55 f^2 + 1.5 f a_1 + a_1^2) &\text{otherwise. }
	\end{cases}
        \label{eq:Ljet}
\end{equation}
We assume $\alpha=0.1$ hereafter; $m_9=M_1/(10^9 M_{\odot}$), where $M_1$ is the mass of the primary BH in the pair; $\dot{m} = \dot{M}/ (22 \,  m_9 \,  M_\odot \,  {\rm yr}^{-1})$ is the central accretion rate in units of the Eddington limit; $a_1$ is the spin parameter of the primary BH; $f$ and $g$ are dimensionless quantities regulating the angular velocity and azimuthal magnetic field of the system, which we set to $f = 1$ and $g = 2.3$ following \cite{2001ApJ...548L...9M}. Equation (\ref{eq:Ljet}) distinguishes between different accretion regimes, defined by the Eddington factor $\epsilon_{\rm edd}$. Indeed, while outflows/jets may occur in optically thick, geometrically thin accretion disks (top row), more powerful jets are produced by advection dominated accretion flows or slim disks (bottom row)~\cite{2001ApJ...548L...9M}. The use of Eq.~\eqref{eq:Ljet}, however, should be accepted with a note of caution. Indeed, the jet powers given by that equation also apply to massive BHs
 accreting steadily, while we are interested here in transient counterparts triggered by a binary merger. 
 It seems reasonable to assume, however, that in the presence of a BH merger the jet should experience
some degree of variability on the system's dynamical time (which is typically of the order of several minutes for eLISA sources), as a consequence of the variability of the gas supply, accretion rate and magnetic field induced by the merger event. Therefore, it seems plausible to assume that the transient's power is also given (at least as an order of magnitude) 
by Eq.~\eqref{eq:Ljet}.

The total luminosity that we will then consider in the radio frequencies is given by $L_{\rm radio} = L_{\rm flare} + L_{\rm jet}$. For a detection of a counterpart by SKA we require \cite{2011ApJ...743..136O}
\begin{equation}
  L_{\rm radio} \geq  4\pi d_L^2 F_{\rm min}^{\rm SKA} \,,
  \label{Eq:LFradio}
\end{equation}
$F_{\rm min}^{\rm SKA}$ being the detector's flux limit. Equation (\ref{Eq:LFradio}) assumes that radio waves are emitted isotropically. This is generally not the case, since synchrotron emission due to particle acceleration within the jet should be beamed along the jet, resulting in a loss of sources depending 
on the jet's opening angle. However, collimation also implies a larger flux for a given source 
luminosity (if the jet points toward the Earth), resulting in intrinsically 
fainter sources being observable. We avoid these complications here, 
and take the isotropic radio flux as a simple approximation.

We assume that the bulk of the emission takes place in the SKA band, and write $F_{\rm min}^{SKA} = \nu_{\rm SKA} F_{\nu,{\rm min}}^{\rm SKA}$ with $\nu_{\rm SKA} \simeq 1.4\,{\rm GHz}$. The current goal of SKA-mid in the phase 1 implementation of SKA is a flux limit of $F_{\nu,{\rm min}}^{\rm SKA} \simeq 2\, \mu{\rm Jy}$ on a 0.5 deg$^2$ field of view, assuming an integration time of 10 minutes\footnote{cf.~Table 1 in the October 2015 SKA baseline description document, \url{https://www.skatelescope.org/wp-content/uploads/2014/03/SKA-TEL-SKO-0000308\_SKA1_System\_Baseline\_v2\_DescriptionRev01-part-1-signed.pdf}}. Full SKA should provide an order of magnitude leap in performance \cite{Lazio}, and we therefore assume that it will be possible to reach $F_{\nu,{\rm min}}^{\rm SKA} \simeq 1\, \mu{\rm Jy}$ on approximately 10 deg$^2$, in the same integration time. Therefore, we select all the events that meet the requirement
\begin{equation}
	\left(\frac{L_{\rm radio}}{\rm erg/s}\right) \left(\frac{d_L}{\rm cm}\right)^{-2} \geq 4\pi\, 10^{-18}\, \left(\frac{F_{\nu,{\rm min}}^{\rm SKA}}{\rm\mu Jy}\right) \left(\frac{\nu_{\rm SKA}}{\rm GHz}\right) \,.
	\label{eq:SKA_cond}
\end{equation}

Radio identification alone, however, does not allow the measurement of the source redshift. We therefore need to follow up the source with optical/IR facilities, looking for spectral features useful for redshift determination. The spectrograph MICADO on the extremely large telescope (ELT) will cover the wavelength range 1000-2400nm ($J$ to $K$ band) \cite{2010SPIE.7735E..2AD}. Spectroscopic observations will be possible down to an apparent magnitude of $m_{\rm ELT,sp}=27.2$ for five hours of integration. This will allow the identification of lines such H$\alpha$ (656nm) to $z\approx 2.6$, [OIII] (500nm) to $z\approx 3.8$, and [OII] (373nm) to $z\approx 5.4$. In addition to that, the Lyman break detection will enable precise photometric redshifts down to an apparent magnitude $m_{\rm ELT,ph}=31.3$ in $J$ and $H$ bands. Moreover, NIRCam \cite{2004SPIE.5487..653W} (mounted on JWST) will have comparable performances down to lower wavelengths (600nm) in less than three hours of integration time. We therefore assume an essentially exact (for our purposes) redshift determination for host galaxies with $m_{\rm gal} \leq m_{\rm ELT,sp}$ and a photometric redshift determination with a conservative uncertainty $\Delta{z}=0.03(1+z)$ (see, e.g., \cite{2013ApJ...775...93D,2013A&A...556A..55I}) for host galaxies with $m_{\rm ELT,sp} < m_{\rm gal} \leq m_{\rm ELT,ph}$.

$m_{\rm gal}$ is computed directly from the host galaxy luminosity through an equation analogue to Eq. (\ref{eq:LSST_cond}):   
\begin{equation}
	m_{\rm gal} = 82.5 - \frac{5}{2} \log_{10} \left(\frac{L_k}{3.02} \frac{\rm s}{\rm erg} \right) + 5 \log_{10}\left(\frac{d_L}{\rm pc}\right)\,, 
	\label{eq:ELT_cond}
\end{equation}
where $L_k$ is the galaxy luminosity in the $K$-band (i.e., with bolometric correction already included), and is computed by converting the host total stellar mass (disk plus bulge), accessible thanks to our galaxy formation model, into luminosity by assuming a fiducial mass-to-light ratio $M/L_k=0.03$. The precise number depends on several factors (including source redshift, dust extinction, etc.). However, for young stellar populations at moderate redshift, $M/L_k$ is quite small, falling in the range 0.01-0.05 \cite{2003MNRAS.344.1000B}. A relatively young stellar population has to be expected, since the merger producing the observed MBHB also triggers intense star formation.

\subsection{Counterpart detection}
\label{sec:counter}

\begin{figure}
\begin{center}
	\includegraphics[width=0.7\textwidth]{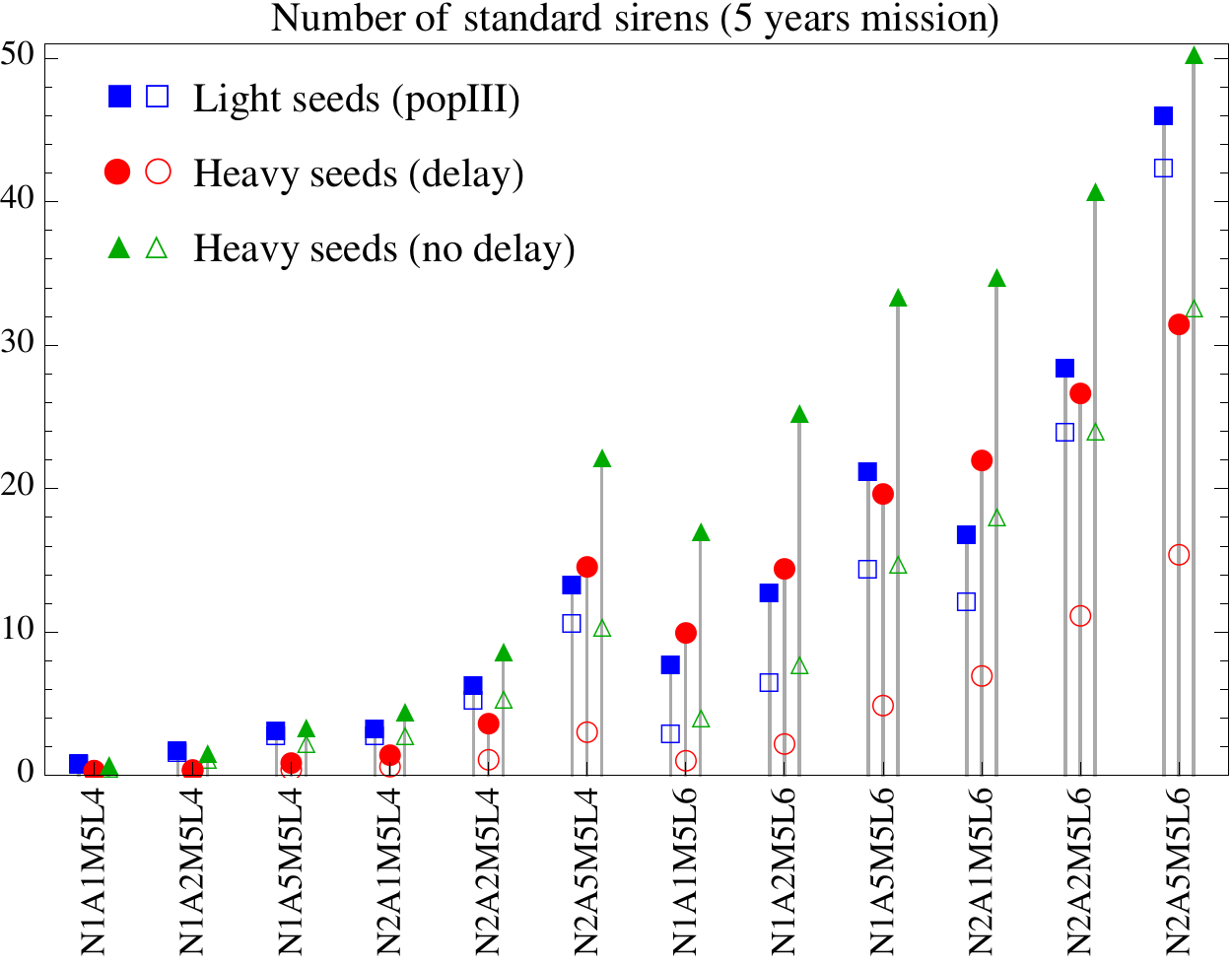}
\end{center}
\caption{Total number of standard sirens available in the conservative (empty markers) 
and optimistic (filled markers) scenarios for each eLISA configuration (corresponding 
to the last columns of Table~\ref{tab:ss} and Table~\ref{tab:ss2}, respectively).}
\label{fig:ss}
\end{figure}

Any electromagnetic counterpart satisfying either Eq.~(\ref{eq:LSST_cond}) or Eqs.~(\ref{eq:SKA_cond}) and (\ref{eq:ELT_cond}) will likely be detectable (and its redshift measurable), and can be used to populate the $d_L-z$ diagram. The number of sources meeting those conditions are reported in Table~\ref{tab:ss} for the conservative scenario and in Table~\ref{tab:ss2} for the optimistic scenario, according to the following structure:
\begin{itemize}
\item For each eLISA configuration, the first row shows the numbers for the light-seed model
 (popIII), the second row for the heavy-seed model with delays (Q3d), and the third row for the 
 heavy-seed model with no delays (Q3nod).
\item The 3rd column shows the number of counterparts directly detected by LSST.
\item The 4th to 6th columns show the number of radio counterparts detected by SKA, respectively from the flare and jet emission and from either of the two.
\item The 7th to 9th columns show the number of SKA detections for which an optical counterpart can be measured by ELT with a five-hour observation. In more detail, the 7th and 8th columns show the number of 
counterparts whose redshift can be measured spectroscopically ($m\leq 27.2=m_{\rm ELT,sp}$) and photometrically ($27.2<m\leq 31.3m_{\rm ELT,ph}$), respectively. The 9th column shows the total detection rate by ELT (sum of the previous two columns).
\item The last column reports the total number of standard sirens available for each configuration (the counterparts being detected either by LSST or by SKA+ELT). These numbers are also displayed in Fig.~\ref{fig:ss}.
\end{itemize}
Note that the numbers in the last two columns of Tables~\ref{tab:ss} and \ref{tab:ss2} always coincide: in fact the counterparts detectable by LSST are always detectable also by SKA+ELT.
Nevertheless, the LSST detections are not completely useless in our analysis:
in fact for some counterparts that are detected by both telescopes, ELT can only measure the redshift photometrically, while LSST is able to measure it spectroscopically.
For these events, the contribution of LSST will thus permit to get rid of the error on the photometric redshift, possibly reducing the overall uncertainty on the determination of the luminosity distance.

From Fig.~\ref{fig:ss} it is clear that in the optimistic scenario the number of standard sirens increases dramatically for configurations with six links, while the situation does not substantially improve for four links, except in the best configurations.
It is also clear that heavy-seed models benefit more than light-seed ones from the addition of the merger and ringdown phases in the search for the counterpart.
This is due to the specific mass distribution in the different models. In the popIII model, most of the eLISA events 
have low mass. For these events the merger and ringdown are not visible by the detector, and adding them to the analysis does not significantly improve the sky localisation. The opposite is true for the Q3d and Q3nod models, for which the addition of the merger and ringdown greatly increases the average SNR of the events, therefore improving sky localisation and ultimately increasing the number of usable standard sirens.
Fig.~\ref{fig:ss} also anticipates what will be confirmed later (see section~\ref{sec:results} and \ref{sec:FoM}) by the Fisher matrix analysis of the cosmological parameters: 
configurations with four links will not provide enough standard sirens to consistently constrain the cosmological parameters.
In fact, among the four-link configurations, only two will detect more than a few standard sirens over a five-year mission period in both scenarios,
namely
N2A5M5L4 and (barely and only in the popIII and Q3nod models) N2A2M5L4.
This might thus prevent an independent cosmological analysis based on a four-link eLISA mission, although the few collected data might turn out useful for integrating other datasets, e.g.~SNIa.
On the other hand, except for the configuration N1A1M5L6 in the conservative scenario, six-link configurations always present a sufficient number of standard sirens to allow for an independent cosmological analysis of the data (see section~\ref{sec:results}), even though Q3d gives systematically fewer counterparts than the other models in the conservative scenario.

\section{Cosmological models}
\label{sec:cosmology}

We consider a Friedmann-Lema\^itre-Robertson-Walker (FLRW) universe with background metric (in polar coordinates and allowing for spatial curvature)
\be
ds^2=-c^2 d t^2 +a^2(t)\left(\frac{dr^2}{1-k\,r^2}+r^2d\Omega\right)\qquad\text{with}\quad k=0, 1, -1 \,.
\ee
From the Friedmann equations, the Hubble rate $H=\dot a/a$ in the late universe can be expressed in terms 
of the redshift $z=a_0/a-1$ as (see e.g.~\cite{ruth})
\begin{equation}
	H(z) = H_0 \sqrt{\Omega_M \left(z+1\right)^3 + \left(1-\Omega_\Lambda -\Omega_M\right) \left(z+1\right)^2
	  +\Omega_\Lambda\, \exp\left[-\frac{3 w_a z}{z+1}\right] \left(z+1\right)^{3 (1+w_0+w_a)}	} \,,
\end{equation}
where $H_0 = h\times 100$ km/(s$\,$Mpc) is the Hubble constant today, and the constituents of the late universe 
are the relative energy density of matter today (dark + baryonic),  
$\Omega_M=8\pi\,G \rho_M^0 /(3H_0^2)$, and the cosmological constant or dark energy energy density today, 
$\Omega_\Lambda=\Lambda c^2/(3H_0^2)$ or $\Omega_\Lambda=8\pi\,G \rho_{DE}^0 /(3H_0^2)$. 
When we consider dark energy, we model its equation of state as $w(z) = w_0 + (1-a) w_a = w_0 + w_a\, z/(z+1)$ \cite{Chevallier:2000qy}, and we therefore have $w_0$ and $w_a$ as extra parameters. In the most general case, we do not assume that the universe is spatially flat: defining an effective relative energy density for the curvature, $\Omega_k=-k c^2/(a_0H_0)^2$, we have $\Omega_k+\OmM+\OmL=1$. The fiducial cosmological model we have adopted has parameter values $\Omega_M = 0.3$, $\Omega_\Lambda = 0.7$, $h = 0.67$ ($H_0 = 67$ km/s/Mpc), $w_0 = -1$, $w_a=0$. 
	
The luminosity distance is $d_L=\sqrt{L/(4\pi\,F)}$, where $L$ is the intrinsic luminosity of a source and $F$ the flux received by the observer. Accounting for the redshift and expansion effects one gets the distance-redshift relation (see e.g.~\cite{ruth,weinberg})
\begin{equation}
	d_L (z) =
	\begin{cases}
		\frac{c}{H_0} \frac{1+z}{\sqrt{|\Omega_k|}} \sin \left[ \sqrt{|\Omega_k|} \int_0^z \frac{H_0}{H(z')} dz' \right] \,, &\text{ if } \Omega_k = 1 - \Omega_M -\Omega_\Lambda >0 \,, \\	
	c \left(1+z\right) \int_0^z \frac{1}{H(z')} dz' \,, &\text{ if } \Omega_k = 1 - \Omega_M -\Omega_\Lambda =0 \,, \\
	\frac{c}{H_0} \frac{1+z}{\sqrt{|\Omega_k|}} \sinh \left[ \sqrt{|\Omega_k|} \int_0^z \frac{H_0}{H(z')} dz' \right] \,, &\text{ if } \Omega_k = 1 - \Omega_M -\Omega_\Lambda <0 \,.
	\end{cases}
\end{equation}
When measuring the distance-redshift relation $d_L(z)$ with observations, one can in principle constrain the values of all the five parameters $(\OmM,\,\OmL,\,h,\,w_0,\,w_a)$. However, there is a strong degeneracy between the parameters $(\OmM,\,\OmL,\,h)$ and the dark energy equation of state parameters $(w_0,\,w_a)$, which makes the simultaneous determination of the five parameters very difficult in practice. Because of this, in the following we analyse the capability of eLISA standard sirens to constrain more specific models of the universe, in analogy with what done e.g.~for the Einstein Telescope in \cite{Zhao:2010sz}, for SNIa in \cite{Betoule:2014frx}, and for BAO in \cite{Aubourg:2014yra}. We consider three subclasses of the full five-parameter model:
\begin{itemize}
\item {\bf $\Lambda$CDM + curvature}: a three-parameter model $(\OmM,\,\OmL,\,h)$, where we fix $w_0=-1$ and $w_a=0$, corresponding to a universe with cosmological constant (CC), dark matter and spatial curvature;
\item {\bf $\Lambda$CDM}: the standard concordance model, where we fix $\OmM+\OmL=1$ and $w_0=-1$, $w_a=0$;
\item {\bf DDE}: a DDE scenario characterised by $(w_0,\,w_a)$, where we fix $\OmM=0.3\,,~\OmL=0.7\,,~h=0.67$.
\end{itemize}
Moreover, we briefly analyse the case where only $\OmM$ and $\OmL$ are taken as free parameters, with the remaining ones being fixed to their fiducial values.
This model is useful to test the curvature and acceleration of the universe, and is mainly considered for comparison with SNIa probes, which are unable to constrain the Hubble constant. We also provide the errors with which eLISA could determine $\OmM$ with all other parameters fixed (to be compared with the SNIa measurement in the $\Lambda$CDM scenario), and $H_0$ with all other parameters fixed (to be compared with cosmic-distance ladder tests, if one assumes the $\Lambda$CDM scenario and fixes the value of $\OmM$).

\section{Statistical method}
\label{sec:statistic}

In order to estimate the errors on the cosmological parameters, 
we adopt a Fisher matrix technique. We simulated a total of 118 five-year catalogues for each MBHB formation scenario 
(popIII, Q3d, Q3nod). This was crucial to get the necessary statistics, since there is a significant scatter in the characteristics of the MBHB population in different catalogues (see Appendix~\ref{app:statistics}).

For each event in each catalogue, we estimate the measurement error on the 
luminosity distance $(\Delta d_L)_n$  for all the twelve eLISA configurations (see section \ref{sec2}), 
by applying a Fisher matrix analysis on the waveforms as
  described in section~\ref{sec:detection}. We then construct the Fisher matrix for the 
  parameters of the cosmological model as
\begin{equation}
		F_{ij}^C = \sum_n \frac{1}{\sigma^2_n + \sigma^2_{\rm lens}(z_n)+ \sigma^2_v(z_n)} \left.\frac{\partial d_L(z_n)}{\partial\theta_i}\right|_{\rm fid} \left.\frac{\partial d_L(z_n)}{\partial\theta_j}\right|_{\rm fid}
		\label{fisher}
\end{equation}
where the sum runs over all MBHB events in a given catalogue, the derivatives of $d_L$ 
(evaluated at the redshift $z_n$ of each event) are performed with respect to the five cosmological parameters
 $\theta_i = (\Omega_M, \Omega_\Lambda, h, w_0, w_a)$ evaluated at their fiducial values, 
 and $C$ identifies the catalogue. To define the typical detector performance,
  we need a suitable combination of the covariance matrices of the 118 catalogues to 
  obtain a unique Fisher/covariance matrix for each MBHB formation scenario and eLISA design.
  This procedure is described at the end of this section, and in more detail in 
Appendix \ref{app:statistics}. Let us here first discuss the uncertainty on the 
luminosity distance, which is given by three terms as shown by 
Eq.~\eqref{fisher}. 

We start with $\sigma_n$, which has several components. First, it contains the eLISA experimental error 
$(\Delta {d_L})_n$, obtained by the Fisher matrix analysis of the waveforms described in section \ref{sec:detection}.
In more detail, the experimental error is determined using both the inspiral and merger and ringdown parts of the waveform, 
for both the conservative and the optimistic scenarios. In fact, whether the counterpart is detected with 
(optimistic scenario) or without (conservative scenario) the help of the merger and ringdown contribution, 
the parameter estimation can be repeated after the merger, when that contribution can always be added.
Moreover, to account for the reduction in this error due to the perfect sky localisation of the system once the counterpart has been detected, we
put an exact prior on the sky position, i.e.~we eliminate the Fisher matrix rows and columns corresponding to the two sky location angles, 
 and then invert the Fisher matrix to obtain the covariance matrix and thus the error on the luminosity distance $(\Delta {d_L})_n$.
Moreover, $\sigma_n$ also includes the error due to the redshift measurement. For all the sources 
whose redshift can be determined  spectroscopically, we assume that this error is negligible for the 
purpose of the present analysis. However, this assumption does not hold for the most distant binaries, 
whose redshift is measured photometrically by ELT. For these sources, 
we estimate the redshift uncertainty approximately 
as $(\Delta z)_n\simeq 0.03(1+z_n)$ 
\cite{2013ApJ...775...93D,2013A&A...556A..55I}. To account for this, 
we propagate this error on the luminosity distance in the standard way, i.e.~we 
assume
\begin{eqnarray}
	 \sigma_n =  \left\{\begin{array}{ll}
	  (\Delta d_L)_n & ~~~{\rm for~spectroscopic~redshift} \,, \\
	  \sqrt{ (\Delta d_L)_n^2 + \left(\frac{\partial d_L}{\partial z} 
(\Delta z)_n\right)^2 } & ~~~{\rm for~photometric~redshift} \,.
\end{array}\right.
\end{eqnarray}

Furthermore, we must account for the systematic error due to weak-lensing. 
Weak-lensing has been estimated as a major source of error on $d_L(z)$ for 
high-redshift standard sirens, degrading the measurement by approximately 5\% 
\cite{Holz:2005df}. In the present analysis, we use the fitting formula of 
\cite{Hirata:2010ba} to estimate the weak lensing error (see also 
\cite{camille}):
\begin{equation}
	\sigma_{\rm lens}(z) = d_L(z) \times 0.066 \left( \frac{1-(1+z)^{-0.25}}{0.25} \right)^{1.8} \,.
	\label{eq:propag_error}
\end{equation}
When we consider the optimistic scenario, we divide $\sigma_{\rm lens}(z)$ by a factor two. The weak lensing magnification uncertainty can in principle be reduced by directly measuring the inhomogeneities in the mass distribution along the line of sight to the source \cite{Dalal:2002wh,Gunnarsson:2005qu,Jonsson:2006vc,2011MNRAS.412.1023H}. Moreover, it has been pointed out that the errors due to weak lensing could be reduced with respect to Eq.~(\ref{eq:propag_error}), 
accounting for the non-Gaussianity of the r.m.s.~of the kappa-distribution \cite{Hirata:2010ba,Shang:2010ta}.
Combining this latter effect with a certain amount of de-lensing, which we assume will be feasible in 2034, we forecast that a de-lensing factor of two
 can be taken into account by the time eLISA is launched. 

Finally the last term in Eq.~\eqref{fisher} is the error due to the peculiar velocities of the GW sources.
In our analysis we use the fitting formula that appears in \cite{Kocsis:2005vv}, namely
\begin{equation}
	\sigma_v(z) = d_L(z) \left[ 1 + \frac{c\ (1+z)}{H(z) d_L(z)} \right] \frac{\sqrt{\left< v^2 \right>}}{c} \,,
	\label{eq:pv_error}
\end{equation}
where $\sqrt{\left< v^2 \right>}$ is the r.m.s.~peculiar velocity of the host galaxy with respect to the Hubble flow, which we fix at 500 km/s, as a rough estimate.

We now discuss how to combine the information of all the catalogues pertaining to a given MBHB formation model. In more detail, we
 aim at quantifying the capability of a given eLISA configuration to constrain cosmology with a unified indicator for each of the cosmological models listed in section \ref{sec:cosmology}. We therefore define a figure of merit (FoM), valid for any cosmological model irrespective of its number of parameters, as
\begin{equation}
	{\rm FoM} = \det(F_{ij}^C)^{\frac{1}{2N}} = \frac{1}{\det(C_{ij}^C)^{\frac{1}{2N}}} \,,
\label{FoM}
\end{equation}
where $C_{ij}^C=(F_{ij}^C)^{-1}$ is the covariance matrix of the $C$th-catalogue, 
and $N$ is the number of parameters of the cosmological model. The FoM  defined in this way 
is the inverse of the $2N$th-root of the volume of the $N$-dimensional parallelepiped whose sides are the 1$\sigma$ 
errors on a suitable combination of the parameters for which there are no correlations. If the errors on the actual cosmological parameters were all equal and uncorrelated, 
the FoM would give the inverse of their value. In the two-parameter case, the above formula is proportional to 
the inverse of the square-root of the area of the ellipse whose axes are given, again, by the 1$\sigma$ error on combinations of the parameters with zero correlation. 
This coincides with the square-root of the FoM defined in the Euclid Study 
Report \cite{Laureijs:2011gra} for the parameters $w_p$ and $w_a$ (see sections \ref{sub:inspiral_only_scenario_} 
and \ref{sub:merger_and_ring_down_scenario}). Note that the FoM as defined by the 
Dark Energy Task Force is different, since it involves the 2$\sigma$ errors of $w_p$ and $w_a$ \cite{Albrecht:2006um}. 

The FoM of the Euclid Study Report conveys consistent information on the errors on $w_p$ and $w_a$, because the two parameters are not degenerate. The FoM defined above in the $N$-parameter case cannot be directly related to the errors on the $N$ parameters, except in the unrealistic case in which these are all exactly equal and uncorrelated. However, it is a useful estimator to compare different eLISA configurations among each other and different cosmological models, irrespective of the number of parameters\footnote{Note that the FoM (\ref{FoM}) does convey some amount of information on the errors, provided that the degeneracies on the cosmological parameters are the same for each eLISA configuration. In this case, by knowing the FoM and the errors on the cosmological parameters for one configuration as well as
 the FoM for another configuration, we could infer the errors on the cosmological parameters given by the second configuration by applying an appropriate scaling.}. 

The FoM in Eq.~\eqref{FoM} is calculated for each of the 118 catalogues of the three MBHB formation scenarios, 
and for each eLISA configuration and each cosmological model. 
The probability distribution function (PDF) of the FoM over the 118 catalogues, for each MBHB formation scenario, 
two eLISA configurations and two cosmological models ($\Lambda$CDM and DDE) 
is shown in Fig.~\ref{fig:PDF}, which will be discussed in Appendix~\ref{app:statistics}, considering both the conservative and optimistic scenario.
In the following, we want to provide a FoM for the capability of a given eLISA configuration to probe a given cosmological model within a given MBHB formation scenario, i.e.~we need to define one single FoM per configuration, cosmological model and MBHB formation scenario. We choose as representative FoM {\it the median of the PDF of the FoM of all the catalogues}. When we compare the different eLISA configurations in section~\ref{sec:FoM}, we will always utilise this median FoM, together with its 10th and 90th percentiles (cf.~Fig.~\ref{fig:FoM}). This will provide an idea of the statistical uncertainty that we can expect in real measurements. 

As explained above, the FoM in Eq.~\eqref{FoM} cannot be directly related to the errors on the cosmological parameters for a given configuration: the
 information on the errors is given by the covariance matrix $C_{ij}$, in particular by (the square root of) its diagonal entries. We thus need to combine the covariance matrices of the 118 catalogues, $C_{ij}^C$, in order to have an estimate of the expected errors of a real measurement with a given eLISA configuration (for a given cosmological model and MBHB formation scenario).
For an estimate of the expected eLISA $1\sigma$ errors on the cosmological parameters, we select 
{\it the median values among the (square roots of the) diagonal entries of the 118 covariance matrices} 
(cf.~Table~\ref{tab:standard_errors}, which will be discussed in section~\ref{sec:results}).
In this estimate, we discard the cases where the low number of data points does not allow to constrain the 
cosmological parameters\footnote{See Appendix~\ref{app:statistics} for more details.}.
In these cases, the expected errors on the parameters are formally infinite, corresponding to useless measurements.
For each eLISA configuration, cosmological model and MBHB formation model, we provide also the percentage $P$ of catalogues for which this does not happen, i.e.~for which it is possible to constrain the cosmological parameters (cf.~Fig~\ref{fig:PoM} in Appendix~\ref{app:statistics}).
This ``probability of measurement'' should be considered as a piece of additional 
information on the ability to constrain a given cosmological model by a given eLISA configuration. 
This is indeed because the standard 1$\sigma$-error estimates provided below are computed by using only the cases where 
fitting of the data is actually possible.
The errors on the cosmological parameters provided in the following should therefore 
be always evaluated {\it in combination with the corresponding probability $P$} (cf.~Table~\ref{tab:standard_errors} and Table~\ref{tab:standard_errors_opt}).

Although an estimate for the standard errors might be defined in this way, in order to plot likelihood contours (``ellipses''), 
we need a consistent Fisher matrix representative of all the 118 catalogues (for a given cosmological model, MBHB formation model, and eLISA design).
We cannot simply take the median of all the covariance matrices entry by entry, 
because some of these entries (i.e., those associated with infinite errors) may be ill-defined, and there is no way to access their information on the degeneracies among the parameters.
For this reason, to combine the covariance matrices $C_{ij}^C$, we have decided to adopt the following criterion: {\it we pick as representative covariance matrix the one that gives the closest FoM to the median of the FoM PDF}. This provides us with well-defined covariance matrices, which can be taken to represent both the estimated median errors and the degeneracies among the cosmological parameters.
Note that instead of taking the median value of the errors, as explained above, we could have used the covariance matrix defined in this way to provide representative errors for each combination of configuration, MBHB formation model and cosmological model.
However this covariance matrix, although well representative of the FoM distribution, still corresponds to a unique realisation of the eLISA data, and is characterised by all its entries and not only by its FoM.
For example, two Fisher matrices with a similar FoM might actually present quite different entries, thus giving quite different errors and degeneracies among the parameters, especially if they are in the tails of the FoM distribution.
For this reason, we have chosen to use the medians of the errors in all catalogues as their representative, 
instead of the errors provided by the covariance matrix closest to the median of the FoM PDF.
In any case, thanks to the large statistics provided by our catalogues, 
we have verified that the two procedures give comparable results (see Appendix~\ref{app:statistics}, in particular~Table~\ref{tab:errors_scatter}).

To have an idea of the possible scatter of the cosmological constraints that we can place,
 we refer to Fig.~\ref{fig:LCDM_percentile} of appendix~\ref{app:statistics}, where we plot the 2$\sigma$ likelihood contours in the $\Lambda$CDM scenario (for the configuration N2A5M5L6 and  the PopIII model) derived from the covariance matrix closest to the median of the FoM PDF, together with those derived from its 20th and 80th percentile.

\section{Constraints on cosmological models}
 \label{sec:results}

We consider the four cosmological models described at the end of section \ref{sec:cosmology} (five-parameter, $\Lambda$CDM with curvature, $\Lambda$CDM, DDE), and for each of them we present results for two representative configurations: N2A5M5L6 (the old LISA) and N2A2M5L4.
The first one gives the best possible result achievable by the eLISA mission (within the considered MBHB models), while the second one represents the worst possible configuration for which independent constraints on the cosmological parameters can be obtained. Worse four-link configurations are unable to constrain cosmology at any level due to the low number of standard sirens (cf.~Fig.~\ref{fig:ss}).
We first discuss the conservative case, in which the sky localisation error is derived using the inspiral phase only. We then move to the optimistic case, in which we also account for merger and ringdown, and we reduce the lensing error by a factor two to mimic de-lensing (see discussion in section \ref{sec:statistic}). The discussion on the FoM and the comparison among all eLISA configurations is presented in section~\ref{sec:FoM}.

\subsection{Conservative scenario (inspiral only)} 
\label{sub:inspiral_only_scenario_}

Table~\ref{tab:standard_errors} summarises the standard $1\sigma$ errors on the cosmological parameters for N2A5M5L6 and N2A2M5L4, for all MBHB formation scenarios and for all cosmological models considered. We also show the errors on $\Omega_M$ and $\Omega_\Lambda$ derived for the case of $\Lambda$CDM with curvature once we fix the Hubble rate to its fiducial value: this is the relevant case to be compared with SNIa (which probe curvature and acceleration), and we discuss it in section \ref{sub:compa}. Moreover, we provide the errors
on $\OmM$, $h$ and $w_0$ once all other parameters but the one under consideration are held fixed: this helps quantify the capability of the given eLISA configuration to constrain these single parameters with respect to other probes; again see section \ref{sub:compa}.

\begin{table}
\hspace{-0.5cm}
\begin{center}
\scalebox{0.85}{\centerline{\begin{tabular}{|c||c|c|c|c|c|c||c|c|c|c|c|c|}
\hline
Model & \multicolumn{6}{|c||}{N2A5M5L6} & \multicolumn{6}{|c|}{N2A2M5L4} \\
\hline
\text{} & $P$(\%) & $\Delta\Omega _M$ & $\Delta\Omega _{\Lambda }$ & $\Delta h$ & $\Delta w_0$ & $\Delta w_a$ & $P$(\%) & $\Delta\Omega _M$ & $\Delta\Omega _{\Lambda }$ & $\Delta h$ & $\Delta w_0$ & $\Delta w_a$ \\
\hline
 \multirow{2}{*}{5} & 100 & 4.31 & 7.16 & 1.58 & 13.2 & 92.3 & 67.8 & 320 & 799 & 47.7 & 344 & 5530 \\
 \multirow{2}{*}{param.} & 100 & 18.0 & 24.9 & 9.95 & 88.6 & 392 & 2.54 & $\gg 10^4$ & $\gg 10^4$ & $\gg 10^4$ & $\gg 10^4$ & $\gg 10^4$ \\
 & 100 & 2.80 & 5.15 & 0.681 & 4.66 & 55.7 & 68.6 & 138 & 306 & 13.3 & 127 & 2400 \\
\hline
 \multirow{2}{*}{$\Lambda$CDM} & 100 & 0.0819 & 0.281 & 0.0521 & \text{} & \text{} & 91.5 & 0.471 & 2.66 & 0.429 & \text{} & \text{} \\
 \multirow{2}{*}{+ curv.} & 100 & 0.220 & 0.541 & 0.136 & \text{} & \text{} & 12.7 & $\gg 10^4$ & $\gg 10^4$ & $\gg 10^4$ & \text{} & \text{} \\
 & 100 & 0.0473 & 0.207 & 0.0316 & \text{} & \text{} & 90.7 & 0.174 & 1.26 & 0.145 & \text{} & \text{} \\
\hline
 & 100 & 0.0473 & 0.0473 & 0.0210 & \text{} & \text{} & 97.5 & 0.275 & 0.275 & 0.0910 & \text{} & \text{} \\
 \text{$\Lambda$CDM} & 100 & 0.0917 & 0.0917 & 0.0480 & \text{} & \text{} & 32.2 & 0.543 & 0.543 & 0.220 & \text{} & \text{} \\
 & 100 & 0.0371 & 0.0371 & 0.0146 & \text{} & \text{} & 99.2 & 0.126 & 0.126 & 0.0400 & \text{} & \text{} \\
\hline
 & 100 & \text{} & \text{} & \text{} & 0.253 & 1.32 & 97.5 & \text{} & \text{} & \text{} & 1.03 & 6.36 \\
 \text{DDE} & 100 & \text{} & \text{} & \text{} & 0.584 & 2.78 & 37.3 & \text{} & \text{} & \text{} & 4.96 & 26.1 \\
 & 100 & \text{} & \text{} & \text{} & 0.176 & 1.00 & 95.8 & \text{} & \text{} & \text{} & 0.427 & 2.87 \\
\hline
 \text{Accel.} & 100 & 0.0190 & 0.0735 & \text{} & \text{} & \text{} & 99.2 & 0.211 & 0.396 & \text{} & \text{} & \text{} \\
 \& curv. & 100 & 0.0280 & 0.105 & \text{} & \text{} & \text{} & 37.3 & 0.977 & 1.30 & \text{} & \text{} & \text{} \\
 test & 100 & 0.0213 & 0.0631 & \text{} & \text{} & \text{} & 94.1 & 0.116 & 0.202 & \text{} & \text{} & \text{} \\
\hline
 \multirow{2}{*}{Error} & 100 & 0.0173 & \text{} & \text{} & \text{} & \text{} & 100 & 0.0670 & \text{} & \text{} & \text{} & \text{} \\
 \multirow{2}{*}{on $\Omega_M$} & 100 & 0.0238 & \text{} & \text{} & \text{} & \text{} & 53.4 & 0.0755 & \text{} & \text{} & \text{} & \text{} \\
 & 100 & 0.0172 & \text{} & \text{} & \text{} & \text{} & 100 & 0.0437 & \text{} & \text{} & \text{} & \text{} \\
\hline
 \multirow{2}{*}{Error} & 100 & \text{} & \text{} & 0.00712 & \text{} & \text{} & 100 & \text{} & \text{} & 0.0146 & \text{} & \text{} \\
 \multirow{2}{*}{on $h$} & 100 & \text{} & \text{} & 0.00996 & \text{} & \text{} & 53.4 & \text{} & \text{} & 0.0175 & \text{} & \text{} \\
 & 100 & \text{} & \text{} & 0.00531 & \text{} & \text{} & 100 & \text{} & \text{} & 0.00853 & \text{} & \text{} \\
\hline
 \multirow{2}{*}{Error} & 100 & \text{} & \text{} & \text{} & 0.0590 & \text{} & 100 & \text{} & \text{} & \text{} & 0.121 & \text{} \\
 \multirow{2}{*}{on $w_0$} & 100 & \text{} & \text{} & \text{} & 0.0786 & \text{} & 53.4 & \text{} & \text{} & \text{} & 0.146 & \text{} \\
 & 100 & \text{} & \text{} & \text{} & 0.0467 & \text{} & 100 & \text{} & \text{} & \text{} & 0.0734 & \text{} \\
\hline
\end{tabular}}}
\end{center}
\caption{Standard 1$\sigma$ errors on the cosmological parameters in the conservative scenario (inspiral only), for two eLISA configurations and in the four cosmological models considered (five-parameter, $\Lambda$CDM with curvature, $\Lambda$CDM, DDE) plus the case $\Lambda$CDM with curvature in which the Hubble parameter $h$ has been fixed to its fiducial value. We add this case here since it provides the relevant tests of acceleration and curvature to be compared with SNIa; cf.~Fig.~\ref{fig:SNe_like} and the associated discussion in section~\ref{sub:compa}. We also add the errors on $\Omega_M$, $h$ and $w_0$ alone, in order to compare them with SNIa and CMB constraints (section~\ref{sub:compa}). The fiducial values are $\Omega_M = 0.3$, $\Omega_\Lambda = 0.7$, $h = 0.67$ ($H_0 = 67$ km/s/Mpc), $w_0 = -1$, $w_a=0$. In each row of the table, the top sub-row shows the errors for light seeds (popIII), the central sub-row for heavy seeds with delays (Q3d) and the bottom sub-row for heavy seeds without delays (Q3nod). Blank entries mean that the corresponding parameter has been fixed to its fiducial value (exact prior). To get the errors in the $\Lambda$CDM model, we impose the flatness constraint $\Omega_M+\Omega_\Lambda =1$, and consequently $\Delta \Omega_\Lambda$ corresponds to $\Delta \Omega_M$. The probability $P$ corresponds to the percentage of catalogues with which it is actually possible to measure the parameters.
The errors in the table are indeed given only by the median among the catalogues providing a measurement, while in the other cases one obtains infinite errors (see the discussion in section~\ref{sec:statistic} and in appendix~\ref{app:statistics}).}
\label{tab:standard_errors}
\end{table}

In each entry of Table~\ref{tab:standard_errors}, the top row shows the errors for light seeds (popIII), the central row for heavy seeds with delays (Q3d) and the bottom row for heavy seeds without delays (Q3nod). Q3nod systematically gives worse results than the other two scenarios -- which are roughly comparable -- due to the lower number of detectable standard sirens, as shown in Fig.~\ref{fig:ss}. In the following, whenever we need to restrict to one MBHB formation model, we always choose popIII, since it is the intermediate scenario (as far as the number of standard sirens is concerned) among those we consider. The constraints would be comparable or slightly better for model Q3nod. 

From Table~\ref{tab:standard_errors}, it appears that even the best eLISA configuration N2A5M5L6 (corresponding to old LISA) is incapable of independently constraining the full five-parameter cosmological model, for which  $(\OmM,\,\OmL,\,h,\,w_0,\,w_a)$ are all free to vary simultaneously. Using eLISA alone, we can only get meaningful constraints on cosmological models described by fewer parameters. This  is indeed also true for all present cosmological probes\footnote{The errors on the full five parameters model are huge compared to present constraints obtained from the combination of other cosmological probes. Therefore we can speculate that marginalising and combining with other probes would not help improving the present constraints. However, we have not performed the analysis yet, here we present results for eLISA only.}. We therefore need to reduce the number of cosmological parameters, by fixing one or more of them to their fiducial value: from the point of view of the statistical analysis, this is equivalent to imposing an exact prior on those parameters.

The reduced Fisher matrices for the three-parameter case of $\Lambda$CDM plus curvature are obtained by removing from the $5 \times 5$ Fisher matrices the rows and columns corresponding to the parameters that are being fixed: in this case, the dark energy EoS parameters, which take the values $w_0 = -1$ and $w_a = 0$. Similarly, the reduced Fisher matrices for the DDE model are obtained by removing from the $5 \times 5$ Fisher matrices the rows and columns corresponding to $(\OmM,\,\OmL,\,h)$. On the other hand, for the $\Lambda$CDM scenario we need first to change variables (from $\Omega_\Lambda$ to $\Omega_k = 1- \Omega_M -\Omega_\Lambda$) to impose the flatness constraint $\Omega_M+\Omega_\Lambda =1$ (corresponding to fix $\Omega_k = 0$), and get the Fisher matrices for the 2-parameters $(\OmM,\,\,h)$.

Note that the four-link configuration N2A2M5L4 does not provide useful constraints on any cosmological model. This is because for this configuration the number of standard sirens with ${\rm SNR}>8$ and $\Delta\Omega<10~{\rm deg}^2$ is small in the conservative scenario under analysis here (cf.~Fig.~\ref{fig:ss} and the discussion in section \ref{sec:counter}). The situation will change in the optimistic scenario discussed in section \ref{sub:merger_and_ring_down_scenario}.

\subsubsection*{$\Lambda$CDM plus curvature $(\Omega_M, \Omega_\Lambda, h)$} 

\begin{figure}
\begin{center}
	\includegraphics[width=\textwidth]{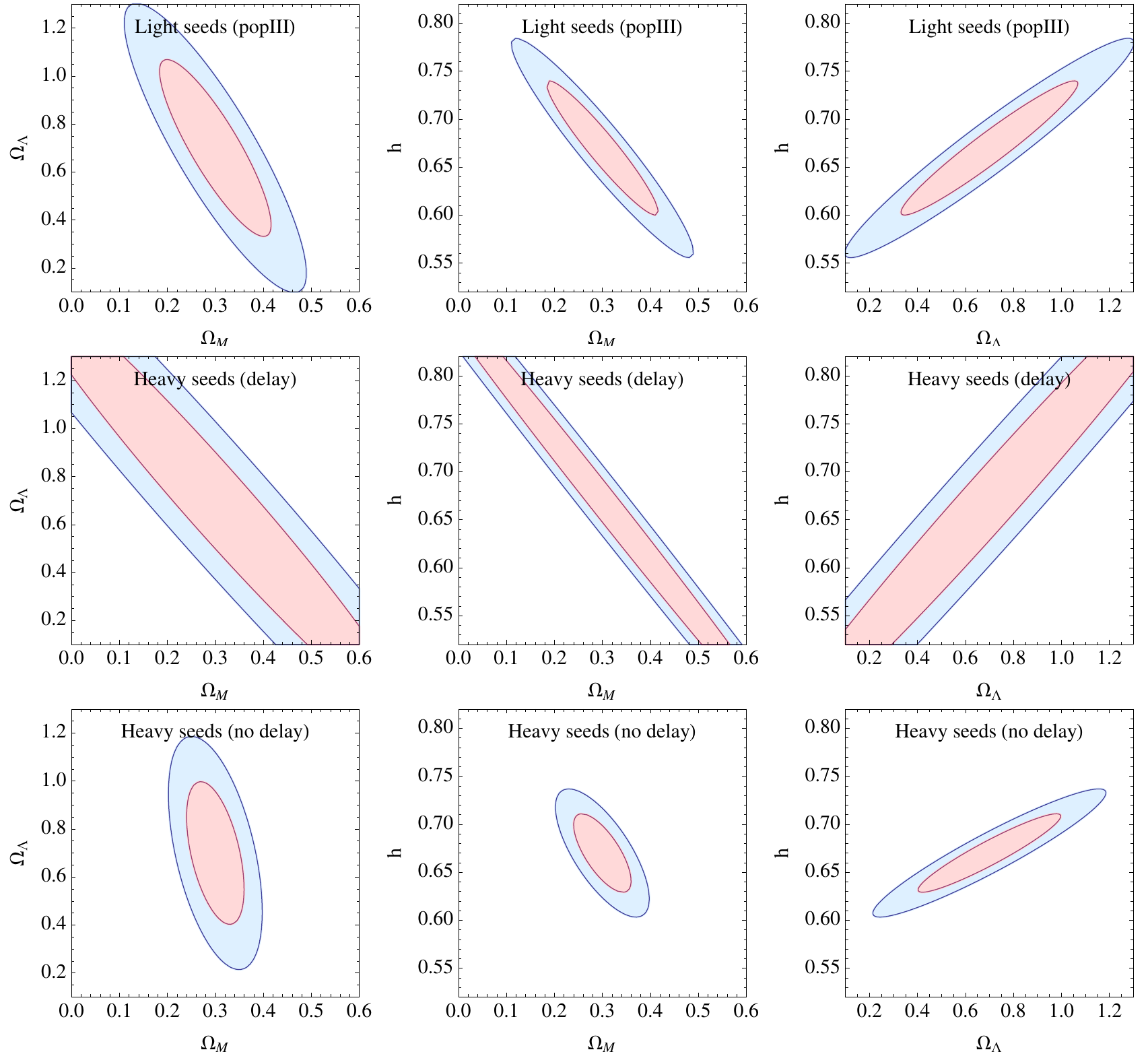}
\end{center}
\caption{Marginalised likelihood contours (1 and 2$\sigma$) for the three-parameter cosmological model ($\Lambda$CDM plus curvature)  in the conservative scenario, with configuration N2A5M5L6. We show all three MBHB formation models. }
\label{fig:Ellipses_3_parameters}
\end{figure}

The first cosmological model that we consider describes a universe with dark matter, dark energy fixed to a CC, and non-vanishing spatial curvature.
The free parameters of the model are $\Omega_M$, $\Omega_\Lambda$ and $h$ and their standard 1$\sigma$ errors can be read from the second row of Table~\ref{tab:standard_errors}, for each MBHB formation scenario.
In Fig.~\ref{fig:Ellipses_3_parameters} we plot the likelihood contours (ellipses) for N2A5M5L6 and for all three MBHB formation scenarios. 
To find the $2 \times 2$ marginalised Fisher matrices used in this plot, we remove from the $3\times 3$ ``representative'' covariance matrix (i.e., the one that gives the closest FoM to the median of the FoM PDF) the row and column of the parameter over which we are marginalising, and then invert the $2\times 2$ matrix obtained in this way to get the new Fisher matrices from which the ellipses are drawn. 

In the $\Lambda$CDM model plus curvature (also called sometimes the $o$CDM model, see e.g.~\cite{Betoule:2014frx}), it is customary to provide the errors on the curvature rather than on $\Omega_\Lambda$. In Table~\ref{tab:Omk} we therefore give the errors on the parameter $\Omega_k$, which are orders of magnitude higher than those obtained by CMB analysis (cf.~for example \cite{Ade:2015xua}). 

It is clear that not even the best eLISA configuration will be able to constrain the cosmological model $\Lambda$CDM plus curvature to a degree comparable with present constraints: these, according to \cite{Betoule:2014frx}, are at the level of 1-3\% by combining SNIa+Planck+BAO. 
On the other hand, once combined with other probes, eLISA in its best configuration may help break the degeneracies among the parameters, in the most favourable scenarios of MBHB formation. eLISA constraints can also naturally complement present electromagnetic measurements, since the redshift range tested by MBHB standard sirens will be much larger (up to $z\sim 8$) and thus affected by different degeneracies in the parameter space.
The combination of eLISA results with other cosmological probes may thus provide new interesting information. 

Moreover, eLISA could give a fully independent constraint on the Hubble parameter, complementary to other direct (optical) measurements. Even though the errors on $h$ are quite high ($\sim 5$\%) compared to other cosmological probes, also in the most favourable MBHB formation models/configurations, 
it is still remarkable that the eLISA mission may potentially measure this parameter, which cannot be independently measured by SNIa observations.
As we will see, the errors on $h$ are even lower for cosmological models with fewer parameters.

\begin{table}
\begin{center}
\begin{tabular}{|c|c|c|}
\hline
 \text{} & N2A5M5L6 & N2A5M5L4 \\
\hline
 \text{} 		  & 0.202 & 2.23 \\
 $\Delta\Omega_k$ & 0.349 & $\gg 10^4$ \\
  \text{} 		  & 0.172 & 1.25 \\
\hline
\end{tabular}
\end{center}
\caption{Errors on the curvature $\Omega_k$ in the three parameter model $\Lambda$CDM model plus curvature, in the conservative scenario. The fiducial value is zero. The top row shows the error for popIII, the central row for Q3d and the bottom row for Q3nod. }
\label{tab:Omk}
\end{table}

\subsubsection*{$\Lambda$CDM $(\Omega_M, h)$} 

\begin{figure}
\begin{center}
	\includegraphics[width=\textwidth]{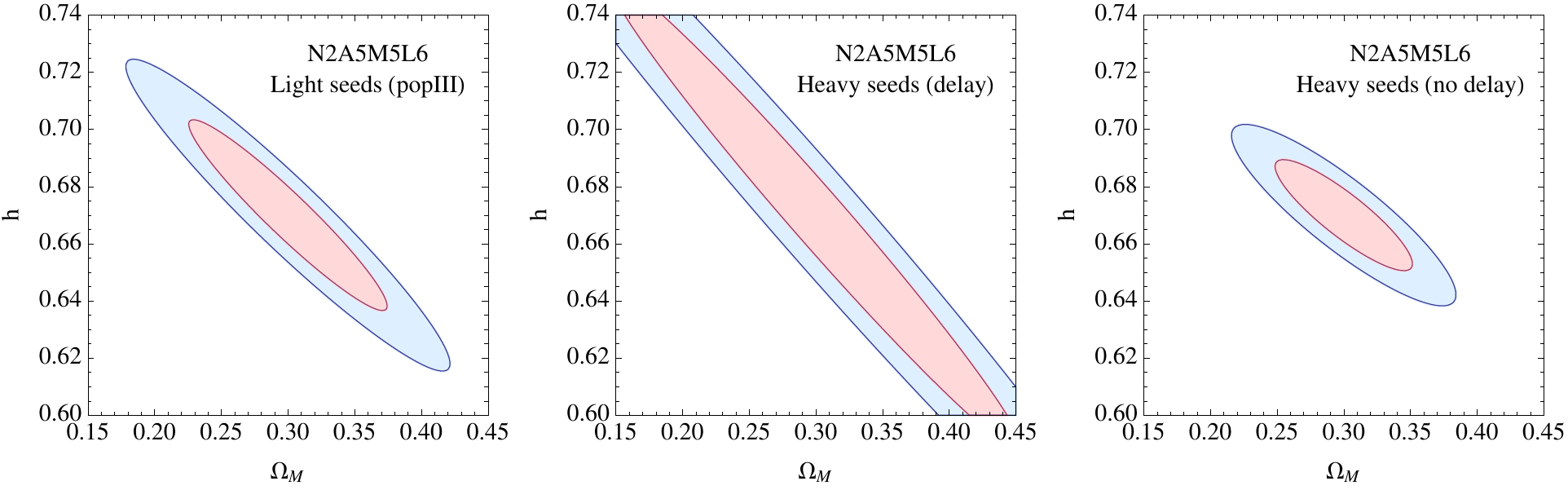}
\end{center}
\caption{1 and 2$\sigma$ likelihood contours for $\Lambda$CDM cosmology in the conservative scenario, with N2A5M5L6, for the three MBHB formation models. We have imposed flatness through $\Omega_\Lambda+ \Omega_M=1$ and fixed $w_0=-1$, $w_a=0$.}
\label{fig:LCDM}
\end{figure}

The standard 1$\sigma$ uncertainties in the $\Lambda$CDM model, characterised by the two parameters $\Omega_M$ and $h$, are given in the third row of Table~\ref{tab:standard_errors}. The $\Lambda$CDM model is evidently much better constrained than the three-parameter one: in the popIII MBHB formation model, N2A5M5L6 gives constraints at the level of 16\% on $\Omega_M$ and 3\% on $h$. 
N2A2M5L4 still does not provide useful constraints, even though the measurement of $h$ starts to become slightly more significant.
Fig.~\ref{fig:LCDM} shows the likelihood contours in the $(\Omega_M, h)$ parameter space for N2A5M5L6. Again, it is apparent that the constraints are much weaker in model Q3d, but in the other two scenarios they are tighter. 

\subsubsection*{DDE $(w_0, w_a)$} 

The ability of any eLISA configuration to constrain DDE is severely limited by the small expected number of MBHBs detected at low redshift (see Fig.~\ref{fig:data} for an example). This is even more true for the conservative scenario under analysis here. 
The best we can expect, using N2A5M5L6 alone, is a FoM as defined by the Euclid Study Report of about 9 (which corresponds to FoM=3 in our definition, cf.~Fig.~\ref{fig:FoM}).
As can be seen from Table~\ref{tab:standard_errors}, in this best situation, eLISA can provide an independent constrain on $w_0$ of 18\%, while the standard 1$\sigma$ deviation on $w_a$ cannot be better than 1.

\begin{figure}
\begin{center}
	\includegraphics[width=\textwidth]{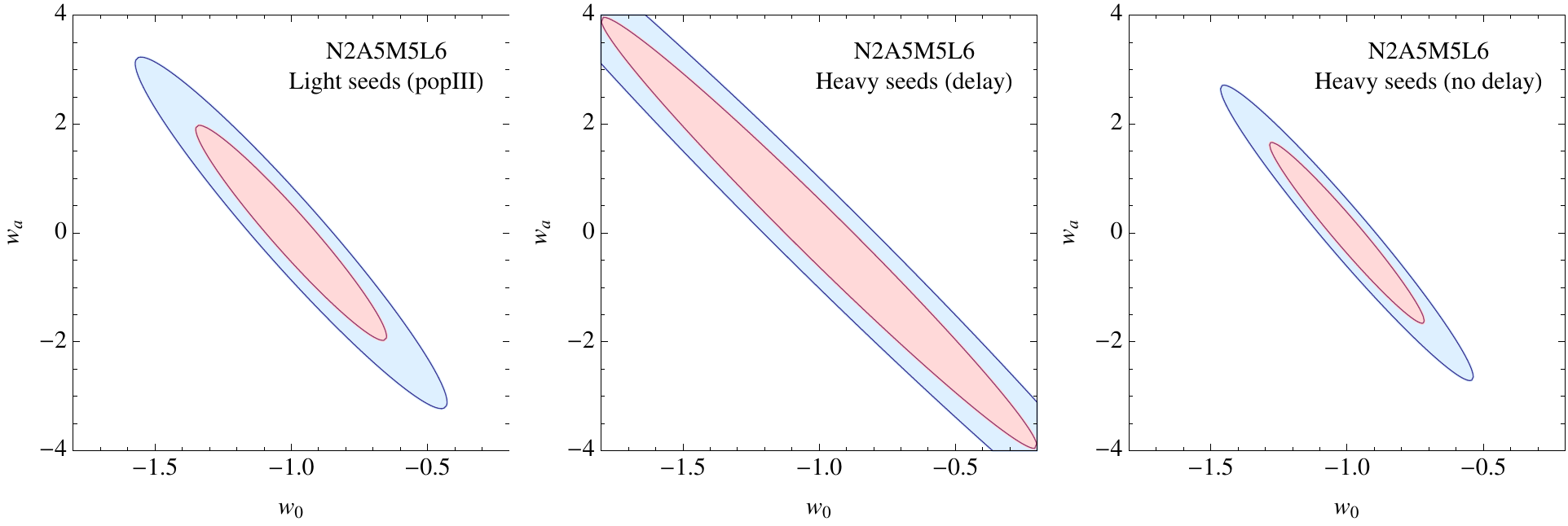}
\end{center}
\caption{1 and 2$\sigma$ likelihood contours for the DDE model $(w_0, w_a)$, in the conservative scenario, for the three MBHB formation models, with N2A5M5L6 (the other cosmological parameters being fixed to their fiducial values).}
\label{fig:DE}
\end{figure}

\begin{figure}
\begin{center}
	\includegraphics[width=\textwidth]{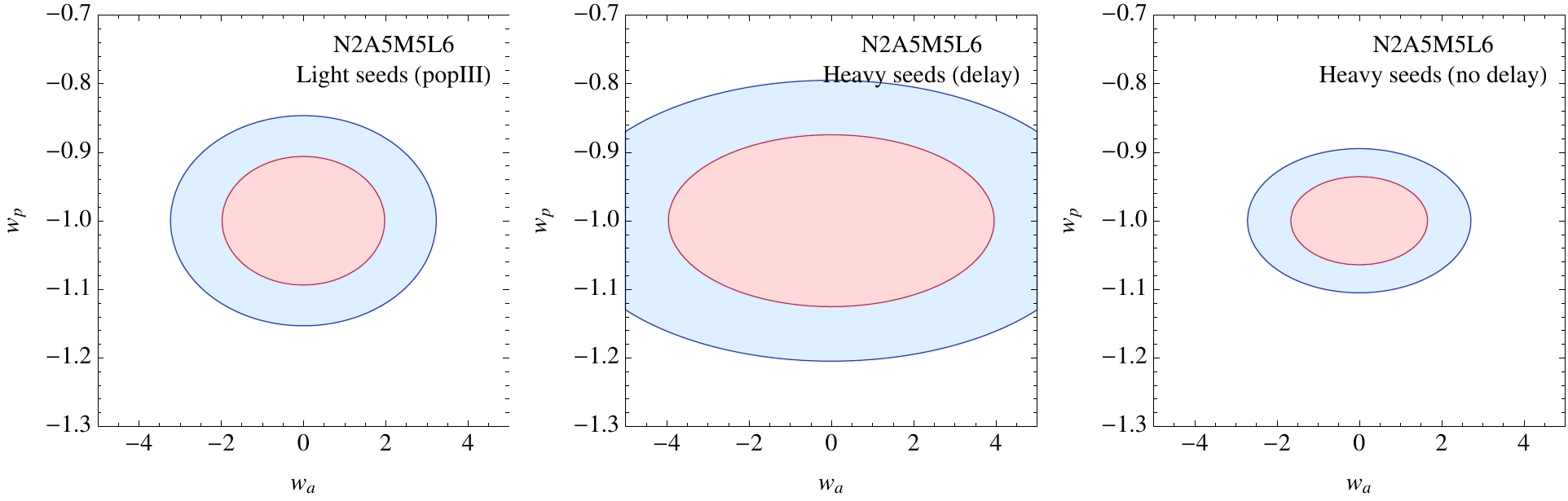}
\end{center}
\caption{1 and 2$\sigma$ likelihood contours for the DDE model with N2A5M5L6 in the $w_p$ and $w_a$ variables, in the conservative scenario, for the three MBHB formation models (the other cosmological parameters being fixed to their fiducial values). }
\label{fig:wp}
\end{figure}

In Fig.~\ref{fig:DE} we have plotted the likelihood contours in the $(w_0, w_a)$ parameter space for N2A5M5L6 (while those for N2A2M5L4 are considerably worse and are not shown).
From this figure, it appears that a strong degeneracy is present between the two parameters $(w_0, w_a)$.
To get rid of this degeneracy, a usual approach is to change variable from $w_0$ to $w_p = w_0 + z_p/(z_p+1)\, w_a$, that is not correlated with $w_a$
for a suitable choice of the pivot redshift $z_p$ \cite{Albrecht:2006um}. Physically, $w_p$ is the DE equation of state parameter at the pivot redshift, i.e.~the redshift at which one can make the best measurement of $w$, defined by $z_p=-[1+\Delta w_a/(\rho \Delta w_0)]^{-1}$ where $\rho$ is the correlation coefficient of $w_0$ and $w_a$. Fig.~\ref{fig:wp} shows the likelihood contours in the uncorrelated variables $(w_p,\,w_a)$. From Table~\ref{tab:pivot} it appears that N2A5M5L6 can measure $w_p$ at the level of 4-6\%, while $\Delta w_a=1$ for the most favourable MBHB formation model\footnote{Note that the errors on $w_a$ given in Table~\ref{tab:pivot} slightly differ from those given in Table~\ref{tab:standard_errors}: this is because they are taken from the covariance matrix used to make the plots in Fig.~\ref{fig:wp} (the one giving the FoM closest to the median of the FoM PDF), while Table~\ref{tab:standard_errors} reports the median of the errors in all catalogues.}.

\begin{table}
\begin{center}
\begin{tabular}{|c|c|c||c|c|c|}
\hline
 \multicolumn{3}{|c||}{N2A5M5L6} & \multicolumn{3}{|c|}{N2A2M5L4} \\
\hline
 $z_p$ & $\Delta w_p$ & $\Delta w_a$ & $z_p$ & $\Delta w_p$ & $\Delta w_a$ \\
\hline
 0.206 & 0.0617 & 1.30 & 0.203 & 0.138 & 5.57 \\
 0.249 & 0.0824 & 2.60 & 0.227 & 0.162 & 32.9 \\
 0.200 & 0.0423 & 1.09 & 0.104 & 0.0881 & 2.29 \\
\hline
\end{tabular}
\end{center}
\caption{Pivot redshift and errors on $(w_p,\,w_a)$  for the conservative scenario. For both eLISA configurations, the top row shows the case of popIII, the central row Q3d and the bottom row Q3nod. }
\label{tab:pivot}
\end{table}

Note that constraining the evolution of $w$ is notoriously difficult. The above quoted errors are in line with those of other present cosmological probes: see for example \cite{Betoule:2014frx} and \cite{Aubourg:2014yra}. In these references the other parameters of the $\Lambda$CDM model ($\Omega_M,\, h$) are not fixed as we do here, but the analysis is done in combination  with the Planck data: this practically amounts to set a very stringent prior on them.
We therefore forecast that eLISA in its best configurations with six links, when combined with other cosmological probes, will be able to help reducing the errors on $w_p$ and $w_a$ in the DDE scenario. The combined analysis will be the subject of future studies, although by the time eLISA is launched, Euclid may already have provided constraints at the level of $\Delta w_a=0.1$ \cite{Laureijs:2011gra}.

\subsection{Optimistic scenario (with merger \& ringdown and de-lensing)} 
\label{sub:merger_and_ring_down_scenario}

Table~\ref{tab:standard_errors_opt} summarises the standard $1\sigma$ errors on the cosmological parameters for N2A5M5L6 and N2A2M5L4, for all MBHB formation scenarios and for all cosmological models considered, exactly as in section \ref{sub:inspiral_only_scenario_} (cf.~Table~\ref{tab:standard_errors}). As before, even the best configuration N2A5M5L6 cannot constrain the full five-parameter cosmological model. Conversely, the situation improves significantly for the other models with respect to the conservative scenario: in particular, the most striking improvement is with configuration N2A2M5L4 in model Q3d. 

\begin{figure}
\begin{center}
	\includegraphics[width=\textwidth]{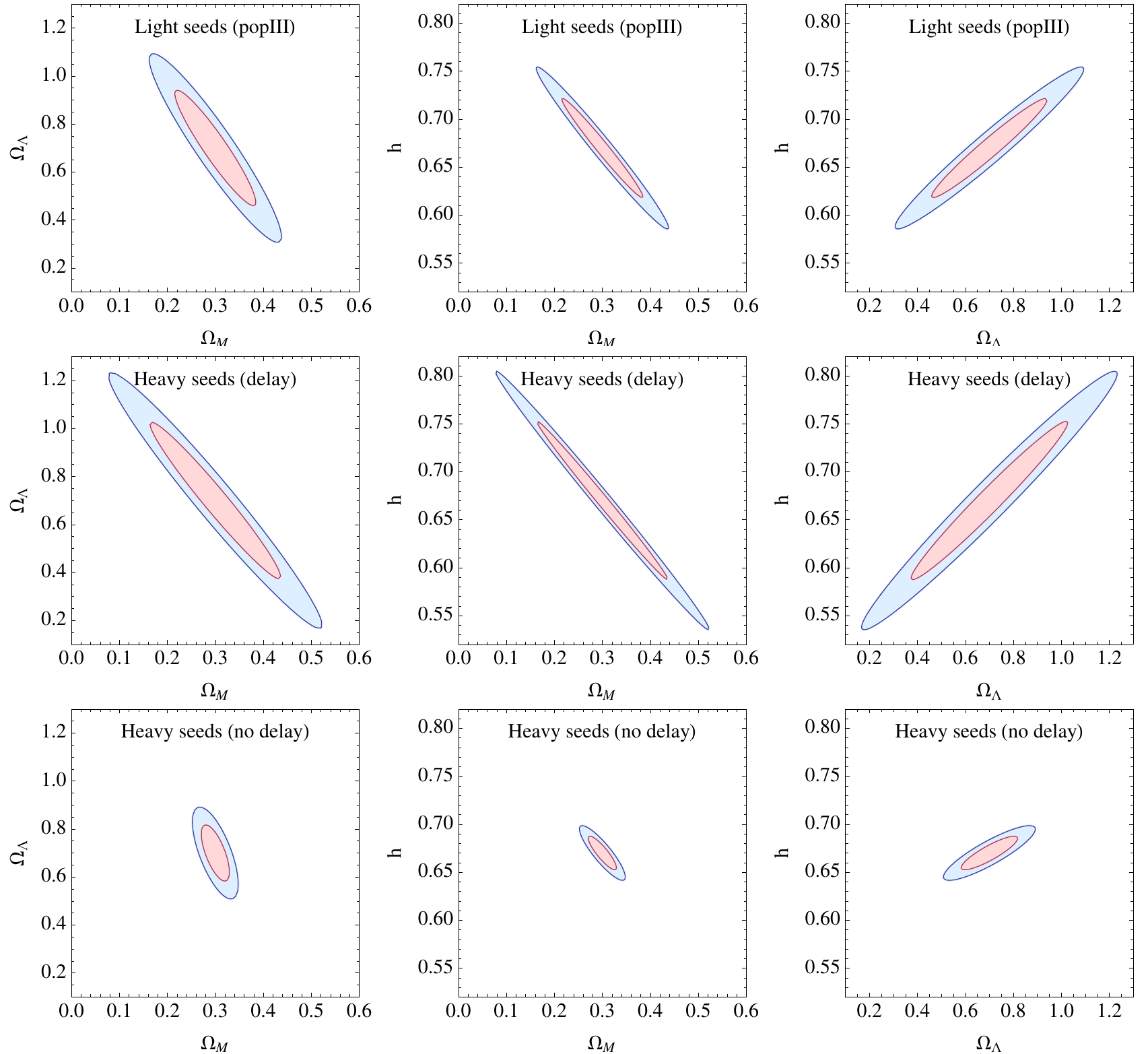}
\end{center}
\caption{Marginalised likelihood contours (1 and 2$\sigma$) for the three-parameter cosmological model ($\Lambda$CDM plus curvature) in the optimistic scenario, with configuration N2A5M5L6. We show all three MBHB formation models. }
\label{fig:Ellipses_3_parameters_opt}
\end{figure}

\begin{figure}
\begin{center}
	\includegraphics[width=\textwidth]{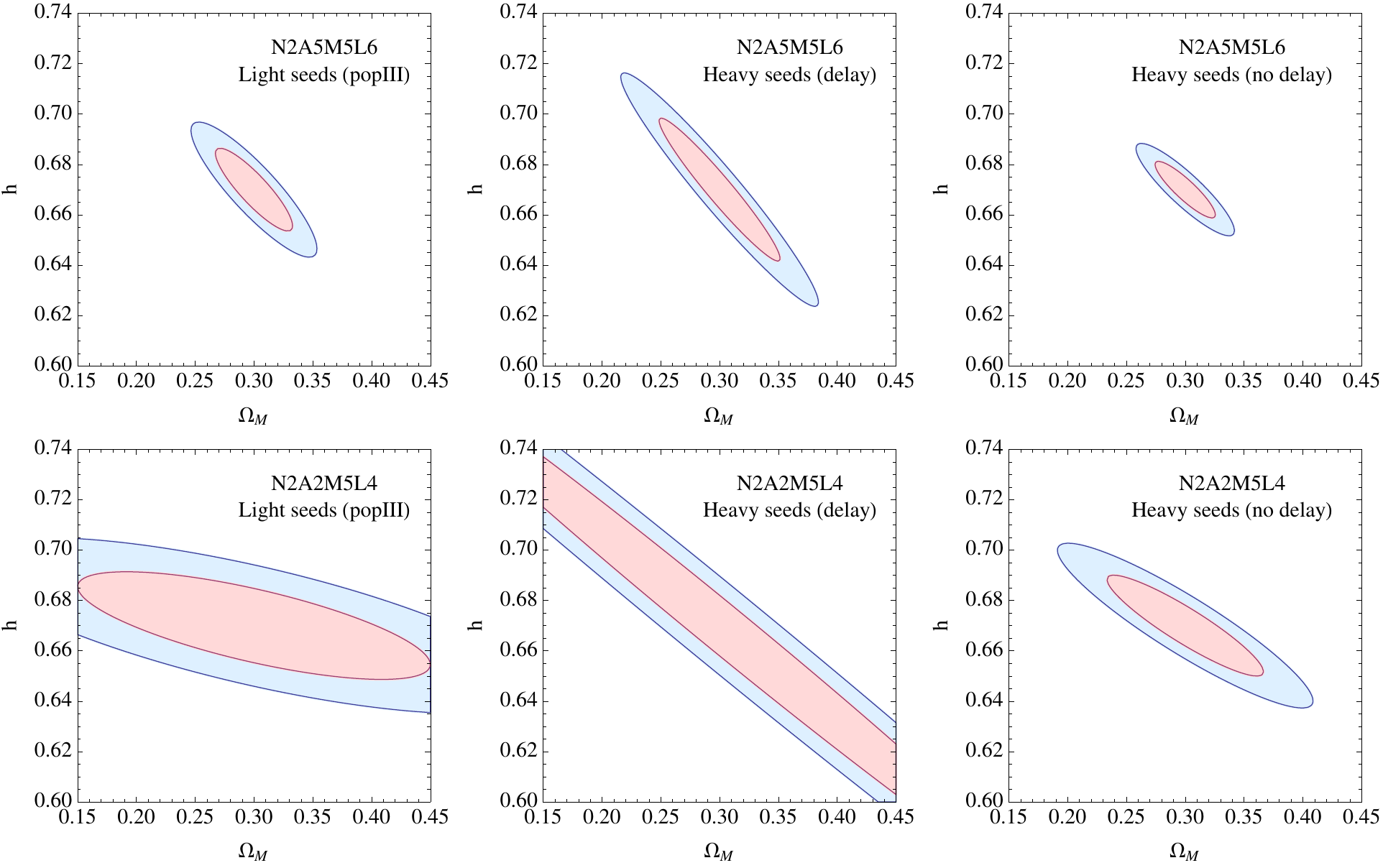} 
\end{center}
\caption{1 and 2$\sigma$ likelihood contours for $\Lambda$CDM cosmology in the optimistic scenario, with both N2A5M5L6 and N2A2M5L4, for the three MBHB formation models. We have imposed flatness through $\Omega_\Lambda+ \Omega_M=1$ and fixed $w_0=-1$, $w_a=0$.}
\label{fig:LCDM_opt}
\end{figure}

In the $\Lambda$CDM plus curvature model, the errors improve by about a factor of two for N2A5M5L6. In particular, $h$ can be measured with a 2.7\% accuracy in the best-case scenario of Q3nod, and this goes down to about 1\% in pure $\Lambda$CDM when $\Om_k= 0$. This level of precision is comparable to the one of present {\it combined} probes SNIa+Planck+BAO \cite{Betoule:2014frx}. The 1 and 2-$\sigma$ contours are shown in Fig.~\ref{fig:Ellipses_3_parameters_opt} for the $\Lambda$CDM plus curvature model (Table~\ref{tab:Omk_opt} shows the errors on $\Omega_k$) and in Fig.~\ref{fig:LCDM_opt} for the $\Lambda$CDM model. Here we show also the contours for N2A2M5L4, since in the optimistic scenario the situation considerably improves for this four-link configuration: in almost 100\% of catalogues, N2A2M5L4 can constrain $\OmM$ to about 50\% and $h$ to about 10\% in the most unfavourable model of MBHB formation, Q3d. These errors are still very large, but compared to the conservative case, in which there was no possibility of measurement, they show {\it the high improvement that can be obtained if a few more standard sirens are available} (thanks to the addition of merger and ringdown in the sky localisation). 

\begin{figure}
\begin{center}
	\includegraphics[width=\textwidth]{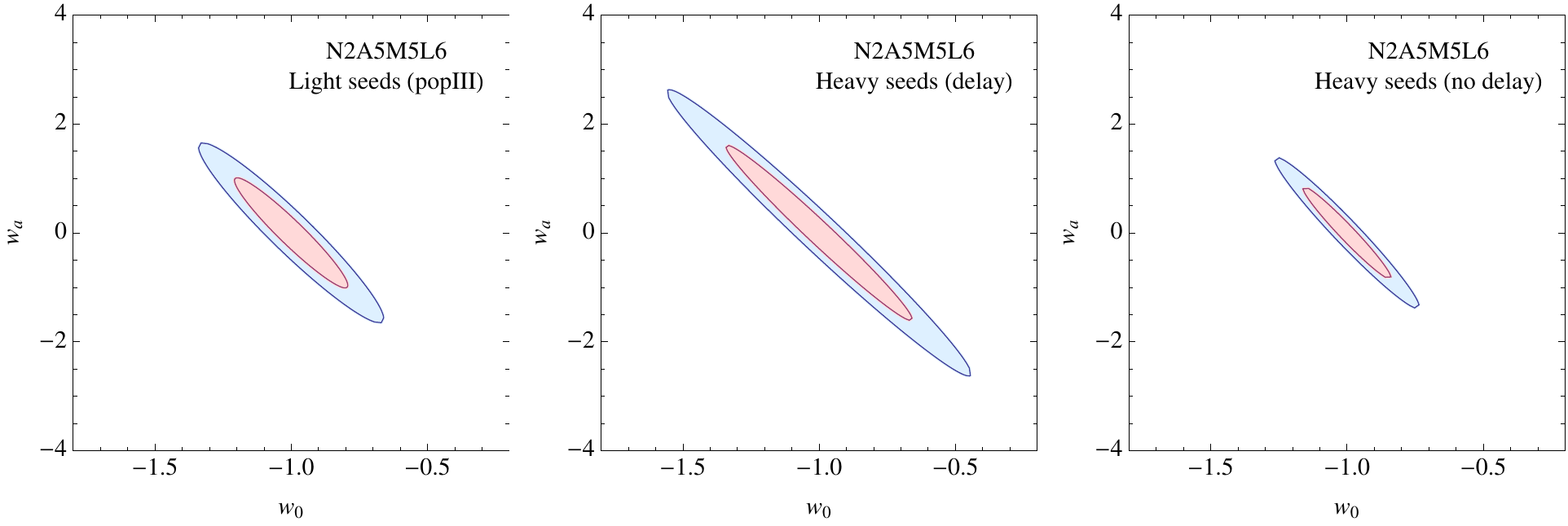}
\end{center}
\caption{1 and 2$\sigma$ likelihood contours for the DDE model ($w_0, w_a$), in the optimistic scenario, for the three MBHB formation models, with N2A5M5L6 (the other cosmological parameters being fixed to their fiducial values).}
\label{fig:DE_opt}
\end{figure}

\begin{figure}
\begin{center}
	\includegraphics[width=\textwidth]{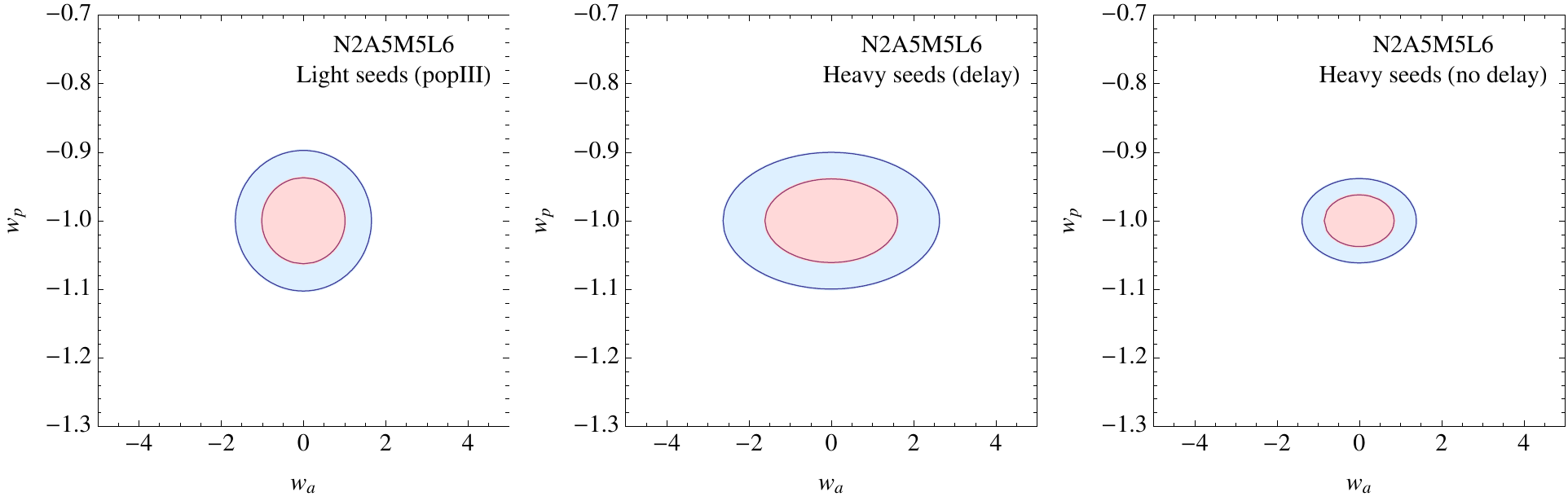}
\end{center}
\caption{1 and 2$\sigma$ likelihood contours for the DDE model with N2A5M5L6 in the $w_p$ and $w_a$ variables, in the optimistic scenario, for the three MBHB formation models (the other cosmological parameters being fixed to their fiducial values). }
\label{fig:wpMR}
\end{figure}

The precision in the measurement of DDE also raises substantially. The 1 and 2-$\sigma$ contours are shown in Fig.~\ref{fig:DE_opt} for $w_0$ and $w_a$, and in Fig.~\ref{fig:wpMR} for the uncorrelated variables $w_p$ and $w_a$ at the pivot redshift. Table~\ref{tab:pivot_opt} gives the errors on $w_p$ and $w_a$.
Again the limitation in constraining the dark energy EoS derives from the low number of data points at low redshift.
Nevertheless, in the optimistic scenario it might be possible to reach FoM$\sim 7$ with the best six-link configurations (cf.~Fig.~\ref{fig:FoM}), corresponding to a FoM as defined by the Euclid Study report of about 50.
From Table~\ref{tab:standard_errors_opt} we see that in the best situation (Q3nod and N2A5M5L6) one can now reach 10\% errors on $w_0$ and a 1$\sigma$ uncertainty on $w_a$ around 0.5, roughly corresponding to a factor 2 improvement with respect to the conservative scenario (cf.~section~\ref{sub:inspiral_only_scenario_}).
Similarly, from Table~\ref{tab:pivot_opt} one can see that N2A5M5L6 might be able to constrain $w_p$ at the level of 2-4\%.
These errors are comparable to the ones obtained by other present cosmological probes (see e.g.~\cite{Betoule:2014frx,Aubourg:2014yra}), meaning that eLISA, at least in its best configurations, may prove useful to reduce the uncertainty on the cosmological parameters if a combined analysis is performed. By the time eLISA is launched, much stronger constraints are expected from Euclid \cite{Laureijs:2011gra}; on the other hand, the eLISA mission will be capable of providing constraints which do not solely depend on EM observations.

\begin{table}
\hspace{-0.5cm}
\begin{center}
\scalebox{0.85}{\centerline{\begin{tabular}{|c||c|c|c|c|c|c||c|c|c|c|c|c|}
\hline
Model & \multicolumn{6}{|c||}{N2A5M5L6} & \multicolumn{6}{|c|}{N2A2M5L4} \\
\hline
\text{} & $P$(\%) & $\Delta\Omega _M$ & $\Delta\Omega _{\Lambda }$ & $\Delta h$ & $\Delta w_0$ & $\Delta w_a$ & $P$(\%) & $\Delta\Omega _M$ & $\Delta\Omega _{\Lambda }$ & $\Delta h$ & $\Delta w_0$ & $\Delta w_a$ \\
\hline
 \multirow{2}{*}{5} & 100 & 2.51 & 4.40 & 0.951 & 8.01 & 55.2 & 80.5 & 120 & 253 & 24.8 & 177 & 2230 \\
 \multirow{2}{*}{param.} & 100 & 4.64 & 6.90 & 2.58 & 22.4 & 103 & 44.1 & 1480 & 3250 & 371 & 2350 & $\gg 10^4$ \\
 & 100 & 1.05 & 1.97 & 0.265 & 2.07 & 21.2 & 93.2 & 12.6 & 27.8 & 2.08 & 15.9 & 227 \\
\hline
 \multirow{2}{*}{$\Lambda$CDM} & 100 & 0.0467 & 0.155 & 0.0299 & \text{} & \text{} & 96.6 & 0.315 & 1.51 & 0.228 & \text{} & \text{} \\
 \multirow{2}{*}{+ curv.} & 100 & 0.0875 & 0.209 & 0.0527 & \text{} & \text{} & 77.1 & 0.396 & 1.61 & 0.306 & \text{} & \text{} \\
 & 100 & 0.0265 & 0.0914 & 0.0161 & \text{} & \text{} & 99.2 & 0.0610 & 0.342 & 0.0520 & \text{} & \text{} \\
\hline
 & 100 & 0.0267 & 0.0267 & 0.0121 & \text{} & \text{} & 99.2 & 0.121 & 0.121 & 0.0445 & \text{} & \text{} \\
 \text{$\Lambda$CDM} & 100 & 0.0368 & 0.0368 & 0.0199 & \text{} & \text{} & 90.7 & 0.151 & 0.151 & 0.0681 & \text{} & \text{} \\
 & 100 & 0.0186 & 0.0186 & 0.00803 & \text{} & \text{} & 100 & 0.0464 & 0.0464 & 0.0159 & \text{} & \text{} \\
\hline
 & 100 & \text{} & \text{} & \text{} & 0.149 & 0.798 & 98.3 & \text{} & \text{} & \text{} & 0.507 & 3.09 \\
 \text{DDE} & 100 & \text{} & \text{} & \text{} & 0.241 & 1.14 & 89.0 & \text{} & \text{} & \text{} & 0.777 & 4.06 \\
 & 100 & \text{} & \text{} & \text{} & 0.101 & 0.544 & 99.2 & \text{} & \text{} & \text{} & 0.201 & 1.20 \\
\hline
 \text{Accel.} & 100 & 0.0105 & 0.0412 & \text{} & \text{} & \text{} & 99.2 & 0.0660 & 0.174 & \text{} & \text{} & \text{} \\
 \& curv. & 100 & 0.00972 & 0.0429 & \text{} & \text{} & \text{} & 84.7 & 0.0544 & 0.161 & \text{} & \text{} & \text{} \\
 test & 100 & 0.00887 & 0.0310 & \text{} & \text{} & \text{} & 99.2 & 0.0381 & 0.0804 & \text{} & \text{} & \text{} \\
\hline
 \multirow{2}{*}{Error} & 100 & 0.00966 & \text{} & \text{} & \text{} & \text{} & 100 & 0.0319 & \text{} & \text{} & \text{} & \text{} \\
 \multirow{2}{*}{on $\Omega_M$} & 100 & 0.00935 & \text{} & \text{} & \text{} & \text{} & 94.1 & 0.0283 & \text{} & \text{} & \text{} & \text{} \\
 & 100 & 0.00788 & \text{} & \text{} & \text{} & \text{} & 100 & 0.0199 & \text{} & \text{} & \text{} & \text{} \\
\hline
 \multirow{2}{*}{Error} & 100 & \text{} & \text{} & 0.00412 & \text{} & \text{} & 100 & \text{} & \text{} & 0.00850 & \text{} & \text{} \\
 \multirow{2}{*}{on $h$} & 100 & \text{} & \text{} & 0.00446 & \text{} & \text{} & 94.1 & \text{} & \text{} & 0.00937 & \text{} & \text{} \\
 & 100 & \text{} & \text{} & 0.00307 & \text{} & \text{} & 100 & \text{} & \text{} & 0.00485 & \text{} & \text{} \\
\hline
 \multirow{2}{*}{Error} & 100 & \text{} & \text{} & \text{} & 0.0342 & \text{} & 100 & \text{} & \text{} & \text{} & 0.0678 & \text{} \\
 \multirow{2}{*}{on $w_0$} & 100 & \text{} & \text{} & \text{} & 0.0368 & \text{} & 94.1 & \text{} & \text{} & \text{} & 0.0729 & \text{} \\
 & 100 & \text{} & \text{} & \text{} & 0.0254 & \text{} & 100 & \text{} & \text{} & \text{} & 0.0416 & \text{} \\
\hline
\end{tabular}}}
\end{center}
\caption{Same as Table~\ref{tab:standard_errors}, but in the optimistic scenario (merger ad ringdown plus de-lensing). }
\label{tab:standard_errors_opt}
\end{table}

\begin{table}
\begin{center}
\begin{tabular}{|c|c|c|}
\hline
 \text{} & N2A5M5L6 & N2A5M5L4 \\
\hline
 \text{} 		  & 0.109  & 1.38 \\
 $\Delta\Omega_k$ & 0.125  & 1.10 \\
  \text{} 		  & 0.0713 & 0.305 \\
\hline
\end{tabular}
\end{center}
\caption{Errors on the curvature $\Omega_k$ in the three parameter model $\Lambda$CDM model plus curvature, for the optimistic scenario. The fiducial value is zero. For both eLISA configurations, the top row shows the error for popIII, the central row forQ3d and the bottom row for Q3nod. }
\label{tab:Omk_opt}
\end{table}

\begin{table}
\begin{center}
\begin{tabular}{|c|c|c||c|c|c|}
\hline
 \multicolumn{3}{|c||}{N2A5M5L6} & \multicolumn{3}{|c|}{N2A2M5L4} \\
\hline
 $z_p$ & $\Delta w_p$ & $\Delta w_a$ & $z_p$ & $\Delta w_p$ & $\Delta w_a$ \\
\hline
 0.244 & 0.0413 & 0.665 & 0.193 & 0.0626 & 3.34 \\
 0.264 & 0.0401 & 1.06 & 0.239 & 0.0830 & 3.30 \\
 0.228 & 0.0248 & 0.559 & 0.194 & 0.0455 & 1.14 \\
\hline
\end{tabular}
\end{center}
\caption{Pivot redshift and errors on $(w_p,\,w_a)$ for the optimistic scenario. For both eLISA configurations, the top row shows the case popIII, the central row the case Q3d and the bottom row the case Q3nod. }
\label{tab:pivot_opt}
\end{table}

\subsection{Comparison with other probes and constraints on single parameters} 
\label{sub:compa}

We end the discussion on the constraints on cosmological parameters by analysing some cases which are instructive for direct comparison with other probes. We use as reference analyses those of \cite{Betoule:2014frx} for SNIa and of \cite{Ade:2015xua} for CMB. 

Let us first analyse the case of $\Lambda$CDM with curvature, once we fix the Hubble rate to its fiducial value; cf.~row five of Table~\ref{tab:standard_errors} and Table~\ref{tab:standard_errors_opt}. This is the usual test of acceleration and curvature, to be compared with the results of SNIa (see for example the discussion in section 6 of \cite{Betoule:2014frx}).
In the conservative scenario, from Table~\ref{tab:standard_errors} it appears that  N2A5M5L6 may constrain $\Omega_M$ at the level of 7-8\% and $\Omega_\Lambda$ at the level of 10-11\% in the popIII and Q3nod models, while in Q3d one obtains slightly worse results. 
N2A2M5L4 worsens these constraints by about one order of magnitude: in the most favourable case (Q3d) one can only achieve uncertainties around 40\% for $\Omega_M$ and around 70\% for $\Omega_\Lambda$, highlighting the difficulties encountered with four-link configurations in the conservative scenario. 
On the other hand, in the optimistic scenario, from Table~\ref{tab:standard_errors_opt} one can notice that N2A5M5L6 manages to constrain $\Omega_M$ at the level of 3\% and $\Omega_\Lambda$ at the level of 5-6\% in all models, while N2A5M5L6 can only reach 13\% for $\Omega_M$ and 11\% for $\Omega_\Lambda$ in Q3nod. These values become respectively 22\% and 25\% for the least favourable MBHB formation model, Q3d.

\begin{figure}
\begin{center}
	\includegraphics[width=.8\textwidth]{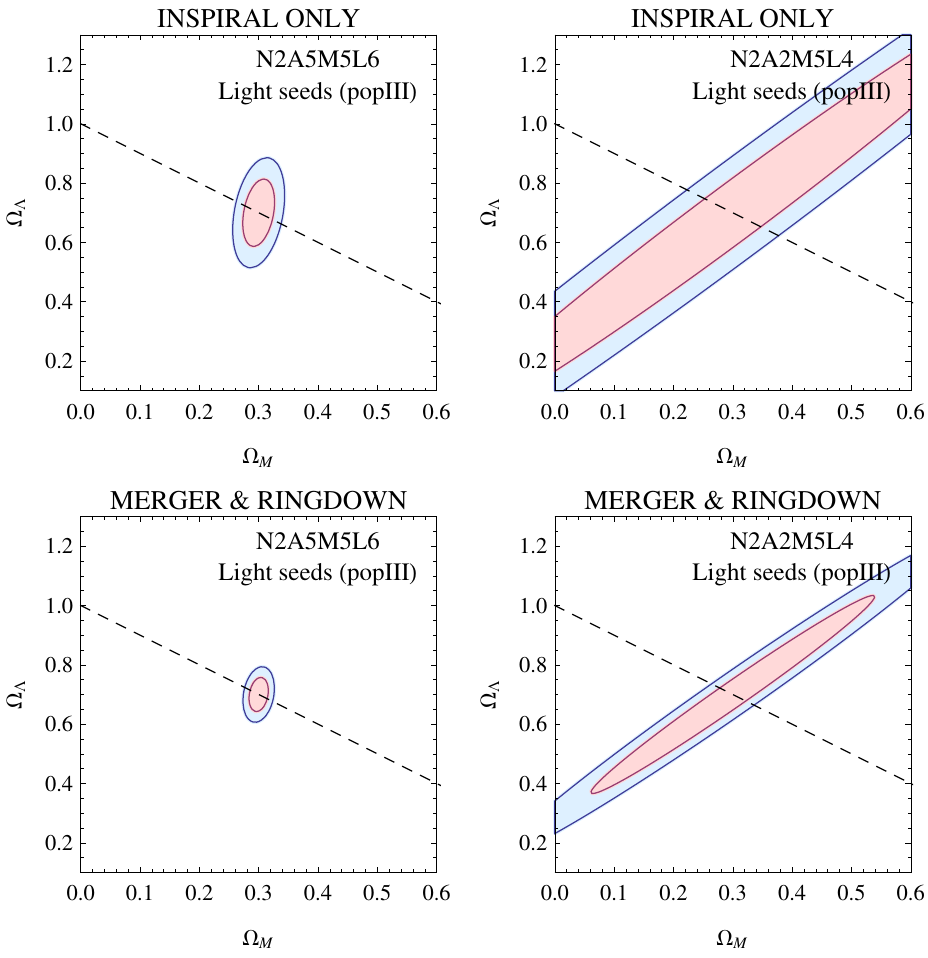}
\end{center}
\caption{1 and 2$\sigma$ likelihood contours in the $\Lambda$CDM with curvature model but with fixed Hubble rate $h=0.67$ (its fiducial value). This contours can be compared directly with the results of SNIa.}
\label{fig:SNe_like}
\end{figure}

If one excludes four-link configurations in the conservative scenario, these errors are competitive to, if not better than, the SNIa results as given e.g.~in \cite{Betoule:2014frx}, especially in the case of the best eLISA configuration and the most favourable MBHB formation models. Likelihood contours for the PopIII model are shown in Fig.~\ref{fig:SNe_like}, for both the conservative and optimistic scenarios.
These contours are much better than the ones obtained with present SNIa data, as one can realise by comparing for example with Fig.~15 of \cite{Betoule:2014frx}.
An important point here is that the degeneracy among $(\Om_M,\,\Om_\Lambda)$ is different from the one of SNIa. Since MBHB standard sirens are visible at higher redshift than SNIa, the likelihood contours for eLISA are less tilted than those for SNIa, as explained in \cite{Goobar:1995af}. Moreover, the degeneracy among these parameters for the CMB combined with BAO measurements is in the opposite direction: cf.~for example Fig.~15 of \cite{Betoule:2014frx}. This means that {\it combining eLISA measurements with SNIa, CMB and BAO will help break the degeneracy in $(\Om_M,\,\Om_\Lambda)$ and will provide even stronger constraints on the acceleration of the universe}.

Another comparison that can be done with SNIa concerns the value of $\Om_M$. Ref.~\cite{Betoule:2014frx} gives the measurement $\Om_M=0.294\pm0.034$ from SNIa only in the $\Lambda$CDM case. This has to be compared with the errors in row six of Table~\ref{tab:standard_errors} and Table~\ref{tab:standard_errors_opt}, i.e.~the $\Lambda$CDM case fixing the Hubble constant to its fiducial value (SNIa in fact do not provide a measurement of the Hubble constant, cf.~section 6 of \cite{Betoule:2014frx}). It appears that {\it eLISA can always constrain $\Om_M$ at a better level than present SNIa}, except in the conservative scenario with N2A2M5L4, where the error is roughly twice the SNIa one. 
More quantitatively, in model Q3nod with N2A5M5L6, one can reach an error of 5.7\% in the conservative scenario and 2.6\% in the optimistic one, thus performing much better than the 11.5\% obtained by present SNIa alone, though not competitive with Planck results \cite{Ade:2015xua}.

Concerning dark energy, the comparison with present observations is discussed at the end of sections \ref{sub:inspiral_only_scenario_} and \ref{sub:merger_and_ring_down_scenario}. Since many cosmological studies, especially previous LISA forecasts \cite{VanDenBroeck:2010fp,Petiteau:2011we}, also consider models in which the EoS of dark energy is a single parameter (without running), in Tables~\ref{tab:standard_errors} and \ref{tab:standard_errors_opt} we have provided the errors that eLISA can obtain once all parameters, except $w_0$, are fixed to their fiducial values. For N2A5M5L6, these are of the order of 5-8\% in the conservative scenario and 2.5-4\% in the optimistic one, in agreement with what found in \cite{Petiteau:2011we}.

Finally, we want to asses the capability of eLISA to constrain the Hubble constant. We consider the $\Lambda$CDM model with $\Om_M$ fixed: the errors are given in row seven of Table~\ref{tab:standard_errors} and Table~\ref{tab:standard_errors_opt}.
In the conservative scenario, N2A5M5L6 can constrain $h$ to 0.8\% in the  Q3nod model and to 1.5\% in the Q3d model, while for N2A2M5L4 the constraints worsen to 1.3\% in Q3nod and to 2.6\% Q3d. On the other hand, for the optimistic scenario, N2A5M5L6 can constrain $h$ up to 0.45\% in Q3nod and to 0.67\% in Q3d, while for N2A2M5L4 the constraints worsen to the range 0.7-1.4\%.

These errors are better than, or at least competitive with, the last CMB measurement (Planck 2015), which is at the level of 1.3\%: from Ref.~\cite{Ade:2015xua}, $h=0.678\pm 0.009$. Note, however, that the quoted Planck measurement is derived by marginalising over all other parameters of the $\Lambda$CDM model (which are six for the CMB), while here we fix them. The marginalised $\Lambda$CDM constraints for eLISA are slightly worse than the CMB one, except in the optimistic scenario for N2A5M5L6; see sections \ref{sub:inspiral_only_scenario_} and \ref{sub:merger_and_ring_down_scenario}.
The relevant point is that {\it eLISA can provide a competitive constraint on $h$, fully independent from other cosmological probes since it does not relies solely on optical measurements}. This will be particularly important given the present tension on the value of $h$ between the CMB, BAO, and cosmic distance ladder data sets (see e.g.~the discussions in \cite{Ade:2015xua} and in \cite{Bennett:2014tka}).

\section{Figures of merit and comparison of eLISA configurations}
\label{sec:FoM}

\begin{figure}
\begin{center}
	\includegraphics[width=\textwidth]{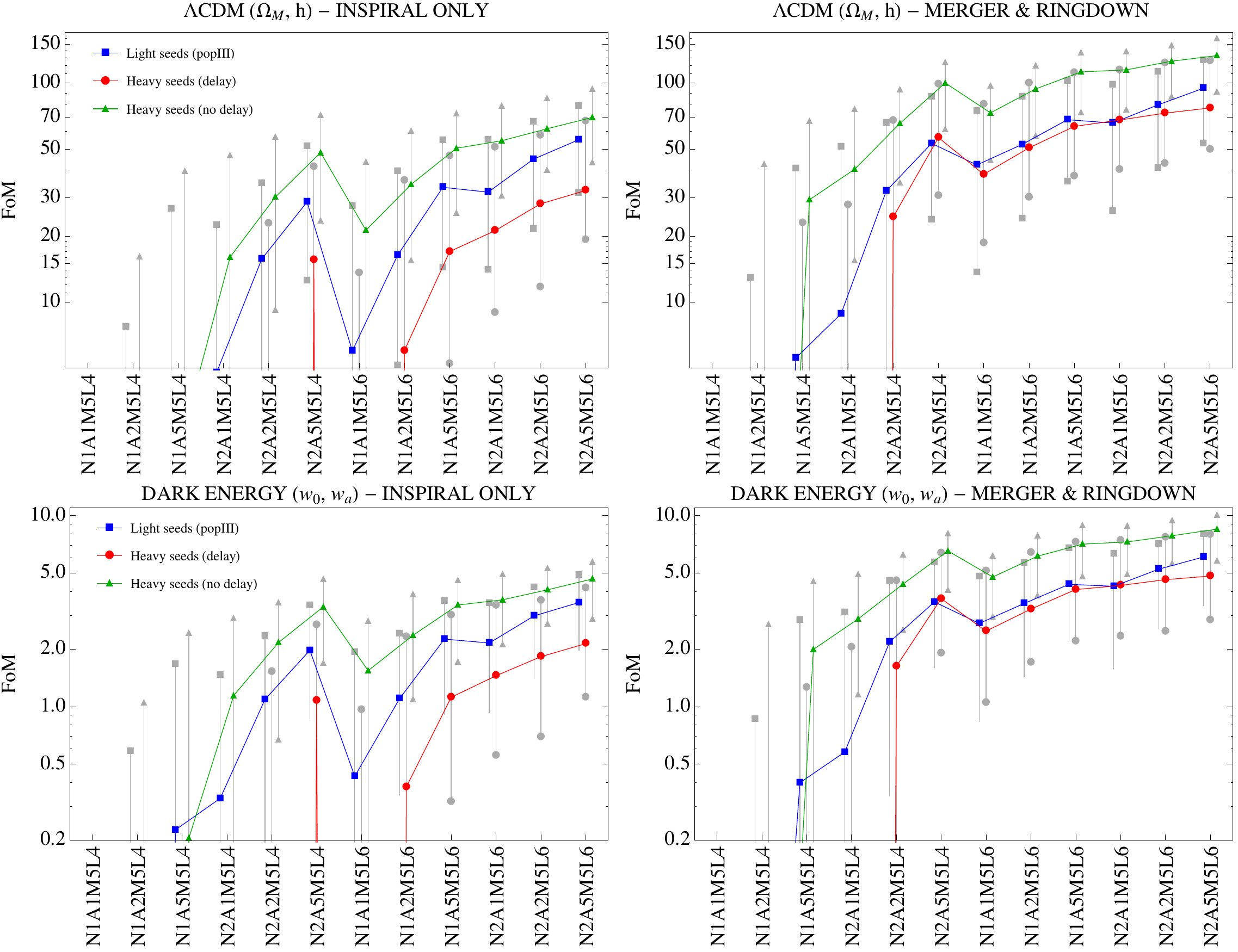}
\end{center}
\caption{Figures of merit as defined in Eq.~\eqref{FoM}, for each eLISA configuration, the three MBHB formation scenarios and two of the considered cosmological models: $\Lambda$CDM and DDE. We show the representative FoM (which corresponds to the median of the PDF of the FoM, see section \ref{sec:statistic}), together with the 10th and 90th percentile (respectively, the lower and upper (grey) points for each configuration). On the left panels we present the FoM for the inspiral only scenario (conservative), while on the right ones we present the FoM for the merger \& ringdown scenario (optimistic).}
\label{fig:FoM}
\end{figure} 

The FoM as defined in section \ref{sec:statistic} can be used as an estimator to compare possible eLISA configurations in their capability to test cosmology. In Fig.~\ref{fig:FoM} we show the FoM for each eLISA configuration, two cosmological models ($\Lambda$CDM and DDE), the three MBHB formation models and both the conservative and optimistic scenarios. As explained in section \ref{sec:statistic}, our simulated catalogues have a considerable scatter: we thus choose as representative the FoM that corresponds to the median of the PDF of the FoMs, but in Fig.~\ref{fig:FoM} we also show the 10th and 90th percentiles to give an idea of the scatter.

The absolute value of the FoMs in Fig.~\ref{fig:FoM} cannot be directly related to the error on the cosmological parameters, because of the well known degeneracies among them. However, their relative values can be used as indicators of the ability of eLISA configurations to constrain different cosmological models.
It is clear from Fig.~\ref{fig:FoM} that $\Lambda$CDM is always better constrained than the DDE model. In fact, this latter model is better tested by sources at low redshift, which are rare for the MBHB standard sirens considered in our analysis. Note that this is the only cosmological model for which the FoM can be directly compared with the errors on $w_p$ and $w_a$, and corresponds to the square root of the FoM defined in the Euclid Study Report \cite{Laureijs:2011gra} (cf.~sections~\ref{sub:inspiral_only_scenario_} and \ref{sub:merger_and_ring_down_scenario}).

Regarding the MBHB formation scenarios, we see from Fig.~\ref{fig:FoM} that in general the Q3nod model systematically gives the best results, while the Q3d model always gives the worse results (although in the optimistic scenario, the results provided by Q3d are comparable with the ones provided by popIII). The reason for this resides in the fact that the three scenarios predict quite different numbers of detectable MBHB events, and therefore of useful standard sirens, as clearly shown in Fig.~\ref{fig:ss} (cf.~section~\ref{sec:counter}).

For what concerns the different eLISA configurations, Fig.~\ref{fig:FoM} shows that the trend does not depend much on the cosmological model. All eLISA configurations with six links, apart from N1A1M5L6 and perhaps N1A2M5L6, have roughly similar FoMs, especially in the optimistic scenario, and therefore have on average the same constraining power on cosmology. In addition, the best configuration with four links (N2A5M5L4) is at the same level as these. Note, however, that configurations with 5 Gm arms are technically very challenging because, for arms longer than 2 Gm, the spacecraft orbits imply a fluctuation of the angles between the arms that requires mechanical pointing of the optical assembly (telescope + optical bench + test mass). It is therefore unfortunate that the only four-link configuration that can give reasonable constraints for all MBHB formation models in the conservative scenario is the one with 5 Gm arms. The results that one can obtain with N2A2M5L4 depend in fact on the MBHB formation model: in Q3d, this configuration has no constraining power, while it improves for the other MBHB formation models. Moreover, in the popIII and Q3nod scenarios, the constraining power of N2A2M5L4 is roughly similar to that of the worst configuration with six links (N1A1M5L6), implying that the errors of the latter are expected to be comparable to the ones presented in Table~\ref{tab:standard_errors} (though slightly worse) and Table~\ref{tab:standard_errors_opt} (though slightly better).
All other L4 configurations give basically no constraints on cosmology: the resulting errors on the parameters are always worse than 100\% and formally infinite in most cases, corresponding to no constraints (cf.~Fig.~\ref{fig:PoM}).
This is mainly due to the low number of standard sirens detected with these configurations (cf.~Fig.~\ref{fig:ss}), as a result of their low sensitivity and, critically, lack of angular resolution.

When we analysed detector performances in the previous section, we focused on N2A5M5L6 and N2A2M5L4, having in mind that they respectively provide the best and worst possible cosmological constraints (other four-link configurations being unable to constrain cosmology at any level).
Fig.~\ref{fig:FoM} confirms this statement, but also shows that in both the conservative and optimistic scenarios, the best four six-link configurations and the best four-link one all provide comparable constraints, being their FoMs approximately similar.
N1A2M5L6 instead yields slightly worse constraints, especially in the Q3d case and in the conservative scenario, while the results obtained with N1A1M5L6 and N2A2M5L6 are even worse.
With all remaining four-link configurations it is impossible to do cosmology.
Focusing on N2A5M5L6 and N2A2M5L4 we therefore provide results that are representative of almost all possible eLISA configurations, as summarised in Table~\ref{tab:eLISA_configurations}. Moreover, the choice of N2A5M5L6 is motivated by the fact that it corresponds to the old LISA configuration (namely the best possible one), while N2A2M5L4 is a reasonable design for a de-scoped mission: as mentioned before, N2A5M5L4 is technically much more challenging and for this reason we preferred not to chose it as a representative four-link configuration.

In the present analysis we have fixed the duration of the mission to five years. We have made this choice partly to reduce the number of free parameters in the analysis, and partly because it seemed to us a very reasonable time duration in perspective. However, in principle the duration is one of the design parameters still under investigation: a two-year mission is also considered as an option at the moment. It is therefore useful to examine how our results change with the mission duration. 
In order to do that, we have reproduced the errors on the cosmological parameters for configuration N2A5M2L6 in the optimistic scenario, by selecting 118 two-year catalogues of MBHBs from our original simulations. This provides a numerical check of the scaling of the errors on the cosmological parameters with the mission duration. The expected behaviour in the absence of correlations is that the errors scale approximatively as $\sqrt{N}$, where $N$ is the number of observed
standard sirens. Since the number of standard sirens is proportional to the mission duration, from five years of observation to two years of observation the errors should increase by a factor $\sqrt{5/2}$. We have verified that, in the one-parameter cosmological models (e.g. all parameters fixed except $H_0$, or except $\Omega_M$, or except $w_0$), this is indeed the case. For cosmological models with two or more free parameters, the loss going from five to two years is instead higher: the errors increase by a factor of two to three for both $\Lambda$CDM and DDE, depending on the MBHB formation model. This is due to the degeneracies between the cosmological parameters, which are usually non-negligible. When we combine the measurements of a five-year catalogue according to Eq.~\eqref{fisher}, slightly different degeneracies concur to give a final combined error volume; the same occurs for a two-year catalogue, but the total number of measured events that concur to give the final combined error volume is smaller, and there are less events with slightly unaligned degeneracies. This effect leads to final errors on the parameters that are larger than one would infer from the simple $\sqrt{N}$ scaling.

More quantitatively, the increase of the errors by a factor of about two to three in the $\Lambda$CDM and DDE cases can be explained as follows. The FoM as defined in Eq.~\eqref{FoM} in the two-parameter case corresponds to the square-root of the inverse of the area of the 1-$\sigma$ likelihood contour (modulo a factor of $\pi$). Going from five to two years, the area of the 1-$\sigma$ ellipse increases by a factor $({\rm FoM}_{({\rm 5y})}/{\rm FoM}_{({\rm 2y})})^2$. Consequently (taking as an example the
 $\Lambda$CDM case),  the errors satisfy the relation 
\begin{equation} 
\Delta\Omega_M^{({\rm 2y})} \Delta H_0^{({\rm 2y})} = \frac{{\rm FoM}_{({\rm 5y})}}{{\rm FoM}_{({\rm 2y})}} \sqrt{\frac{1-\rho_{({\rm 5y})}^2}{1-\rho_{({\rm 2y})}^2}}
\,\Delta\Omega_M^{({\rm 5y})} \Delta H_0^{({\rm 5y})}\,, 
\end{equation}
where $\rho_{({\rm 5y})}$ and $\rho_{({\rm 2y})}$ denote the correlation coefficients in the five and two-year cases respectively. We have verified that the square root of this relation approximately holds also for each error singularly, providing the above mentioned factor two to three (depending on the cosmological and MBHB formation models). For example, we have
\begin{equation}
\Delta\Omega_M^{({\rm 2y})} \simeq \sqrt{\frac{{\rm FoM}_{({\rm 5y})}}{{\rm FoM}_{({\rm 2y})}}} \left(\frac{1-\rho_{({\rm 5y})}^2}{1-\rho_{({\rm 2y})}^2}\right)^{1/4}
\Delta\Omega_M^{({\rm 5y})} = 2.8 \, \Delta\Omega_M^{({\rm 5y})} \,,
\end{equation}
where the second equality holds specifically in the $\Lambda$CDM case for the popIII MBHB formation scenario. 

To summarize, for missions lasting less than five years, the errors in the one-parameter family of cosmological models increase with respect to what found in the present analysis by a factor corresponding to the square root of the ratio between five years and the mission duration. For the two-parameter cosmological models, the increase in the errors is in general higher, as a result of the big degeneracies existing between the cosmological parameters. It is difficult to predict the actual increase factor without recalculating the Fisher matrix, since it depends on its determinant and correlation coefficients. Going from N2A5M5L6 to N2A5M2L6 in the optimistic scenario, the errors increase by a factor of two to three in both the $\Lambda$CDM and DDE cosmological models: we expect the same factor to hold for all configurations with FoMs similar to N2A5M5L6, but to increase substantially for worse configurations (namely N2A1M5L4 and worst), due to the poorer statistics leading to lower FoMs. 

\begin{table}
\begin{center}
\begin{tabular}{|c|c|}
\hline
N2A5M5L6 & \multirow{5}{7.5cm}{Constraints comparable to or slightly worse than N2A5M5L6 (cf.~Table~\ref{tab:standard_errors} and Table~\ref{tab:standard_errors_opt}).} \\
N2A2M5L6 & \\
N2A1M5L6 & \\
N1A5M5L6 & \\
N2A5M5L4 & \\
\hline
\multirow{2}{*}{N1A2M5L6} & \multirow{1}{7.5cm}{Constraints worse than N2A5M5L6, but better than N2A2M5L4.} \\
& \\
\hline
N1A1M5L6 & \multirow{2}{7.5cm}{Constraints comparable to or slightly better than N2A2M5L4 (cf.~Table~\ref{tab:standard_errors} and Table~\ref{tab:standard_errors_opt}).} \\
N2A2M5L4 & \\
\hline
N2A1M5L4 & \multirow{4}{7.5cm}{Constraints worse than N2A2M5L4 or no constraints at all.} \\
N1A5M5L4 & \\
N1A2M5L4 & \\
N1A1M5L4 & \\
\hline
\end{tabular}
\end{center}
\caption{Comparison of different eLISA configurations in constraining cosmology (see section~\ref{sec:FoM}).}
\label{tab:eLISA_configurations}
\end{table}

\section{Conclusions}
\label{sec:conclusions}

This work has been devoted to the investigation of one of the possible cosmological implications of the eLISA mission, namely the constraints that can be put on the cosmological parameters by using GW standard sirens. Considering three models for the MBHB formation (heavy seeds without delays, heavy seeds with delays and popIII stars), we have produced catalogues of MBHB events and selected those for which a counterpart may realistically be observed, according to the procedure outlined in section~\ref{sec:counterparts}. The main issue for the identification of a counterpart resides in the capability of obtaining a sufficiently accurate sky localisation: in this respect, we have considered both a conservative and an optimistic scenario. In the first one the sky localisation error is assumed to be determined only by the analysis of the inspiral waveforms; in the second one, by 
analysing the whole waveforms including also the merger and ringdown phases. In this latter scenario we have also assumed a de-lensing by a factor of two (see discussion in section \ref{sec:statistic}). 

If only the inspiral phase can be used to localise the counterpart, fewer standard siren events are available to constrain cosmology. In this case, we have found that meaningful constraints can be provided only by the six-link configurations with best noise level (N2), or by N1A5M5L6 and N2A5M5L4. This is true in all the three MBHB formation models, though Q3d gives slightly worse results. For all cosmological models considered, the constraints on the parameters by eLISA alone are systematically worse than the current ones, obtained by combining different cosmological probes (SNIa, CMB and BAO). However, even in this conservative scenario, eLISA can be used to set an independent (i.e., not relying exclusively on electromagnetic measurements) constraint on the Hubble parameter at the level of 1-2\%, depending on the mission configuration (with all other parameters fixed to their best fit value). On the other hand, due to the small number of low redshift events, the constraints in the conservative scenario on DDE are not competitive with present combined constraints, and will therefore not be relevant by the time eLISA is launched, i.e.~after Euclid.  

If the error on the sky location can be reduced by including the merger and ringdown phases, and a de-lensing by a factor two can be taken into account, the power of eLISA to constrain cosmological parameters considerably improves. All six-link configurations can provide significant constraints, together with the best two four-link configurations (N2A5M5L4 and N2A2M5L4). Therefore, besides improving the overall constraining power of the configurations that were relevant in the conservative scenario, one also gains three other potentially useful configurations: N1A2M5L6, N1A1M5L6 and N2A2M5L4. The measurement of the Hubble parameter can reach a precision of 1-3\% depending on the MBHB formation model in the $\Lambda$CDM scenario, i.e.~without fixing $\Omega_M$, for all six-link configurations with the best noise level, plus N1A5M5L6 and N2A5M5L4. This goes down to 0.45-0.7\% if one fixes $\Omega_M$ to its fiducial value, providing therefore a measurement of $h$ far more precise than current constraints. For the remaining interesting configurations (N1A2M5L6, N1A1M5L6 and N2A2M5L4) these errors degrade by a factor $\sim 2$--3.
Moreover, DDE can be probed in the optimistic scenario to a level comparable with present constraints: down to 2-4\% for $w_p$ with $\Delta w_a=$0.5--1, depending on the MBHB formation model, for all six-link configurations with the best noise level, plus N1A5M5L6 and N2A5M5L4. The remaining N1A2M5L6, N1A1M5L6 and N2A2M5L4 configurations give 4-8\% accuracy on $w_p$ and $\Delta w_a=$1--3. 

As for the constraints on $\Omega_M$, even in the best case scenario, namely with the N2A5M5L6 configuration, the Q3nod model, and fixing all other cosmological parameters, one can at most reach a precision of 2.6\%: yet not comparable to the one of CMB measurements (which is at the level of 0.4\%), but better than the one reached by present SNIa catalogues (which is at the level of 11\%).
The constraints in the $(\Omega_M, \Omega_\Lambda)$ parameter space are also better than those provided by SNIa (cf.~Fig.~\ref{fig:SNe_like}), and they are characterised by a different degeneracy, since MBHB standard sirens are visible at higher redshift.
Indeed, the novelty with respect to the CMB and SNIa measurements resides in the fact that eLISA has the potential to probe the expansion of the universe in the redshift range $1\lesssim z\lesssim 8$, which has not yet been tested by present observations. Although in the $\Lambda$CDM model, for this redshift range, the universe is supposed to be simply dominated by dark matter and decelerate accordingly, eLISA could prove useful to constrain alternative cosmological models predicting a different expansion rate at very high redshift: for example, early or interacting dark energy, decaying dark matter, unified dark matter models and so on (see e.g.~\cite{Aubourg:2014yra,Ade:2015rim} where a similar analysis has been done with BAO and CMB data).
The investigation of these alternative cosmological models will be considered in a future study.

It has been pointed out that a well-identified electromagnetic counterpart may not be necessary to measure the cosmological parameters with eLISA: 
a statistical method for the  identification of a source's redshift can be used, as described in \cite{MacLeod:2007jd,Babak:2010ej,Petiteau:2011we}. For eLISA, this method has the advantage of exploiting not only MBHBs as standard sirens, but also extreme mass ratio inspirals (EMRIs), which should be detected in large numbers (see \cite{AmaroSeoane:2012je} and references therein). EMRIs are expected to be more abundant than MBHBs at low redshift, and might therefore provide a better way to probe the late time acceleration and DDE models, improving in this way the constraints obtained in the present analysis. This study is currently under investigation within the eLISA consortium.

In summary, standard sirens with eLISA {\it in its best configurations} have the potential to provide a complementary and calibration-free way to test the expansion of the universe (and in particular $H_0$) with entirely different systematics than present probes (e.g. SNIa), which rely on electromagnetic observations only. This would constitute an important additional scientific gain for a mission that has been designed to detect the gravitational signal of MBHBs.

\acknowledgments

We would like to thank Volker Beckmann, Emanuele Berti, Marc Betoule, Marco Cirelli, Zoltan Haiman, Daniel Holz, Bradley Kavanagh, Joseph Lazio, Camilla Pacifici and Daniela Paoletti for providing us with very valuable informations that were used in this analysis.
E.~Barausse acknowledges support from the European
Union's Seventh Framework Programme (FP7/PEOPLE-2011-CIG) through the
Marie Curie Career Integration Grant GALFORMBHS PCIG11-GA-2012-321608, and
from the H2020-MSCA-RISE-2015 Grant No. StronGrHEP-690904.
A.~Sesana is supported by a University Research Fellowship of the Royal
Society.
A.~Klein is supported by NSF CAREER Grant
No.~PHY-1055103.
N.~Tamanini acknowledge support from the Labex P2IO and the Enhanced Eurotalents Programme.
Computations were
performed on the gpc supercomputer at the SciNet HPC Consortium, as well as on the ``Projet Horizon Cluster'' at the Institut d'astrophysique de Paris.
N.~Tamanini and C.~Caprini would like to thank for hospitality the {\it Institut d'Astrophysique de Paris} and the institute {\it AstroParticule et Cosmologie}, where part of this work has been completed. 

\appendix
\section*{Appendix}
\section{Quasar bolometric luminosity}
\label{app:luminosity}
Our semi-analytical galaxy formation model tracks the evolution of the ``reservoir'' of nuclear gas 
available for accretion onto the central massive BH under a variety of physical processes, represented schematically in 
Fig.~\ref{fig:model}. In more detail, we follow \cite{mymodel,spin_model,letter,newpaper} and assume that accretion takes place on a ``viscous'' timescale
\begin{equation}
t_{\nu}\approx {\rm Re}\, t_{\rm dyn}\,,
\end{equation}
where ${\rm Re}\approx 10^3$ is the critical Reynolds number marking the onset of turbulence, and
$t_{\rm dyn}=G M_{\rm bh}/\sigma^3$ ($M_{\rm bh}$ and $\sigma$ being the BH mass and the stellar velocity dispersion) is the dynamical time at the BH's influence radius.
As such, the mass accretion rate $\dot{M}_{\rm bh,\,QSO}$ onto the massive BH is computed as
\begin{gather}
\dot{M}_{\rm bh,\,QSO}=\min\left(\frac{M_{\rm res}}{t_{ \nu}},A_{\rm Edd} \dot{M}_{\rm Edd}\right)\,,\\
\dot{M}_{\rm Edd}=\frac{L_{\rm Edd}}{\eta(a_{\rm bh}) c^2}\,,
\end{gather}
where $M_{\rm res}$ is the reservoir's mass, $L_{\rm Edd}=1.26\times 10^{38} \left( {M}/{M_\odot} \right)\,{\rm erg/s} $ is the Eddington luminosity, and
$\eta(a_{\rm bh})$ is the BH's radiative efficiency (which is in turn a function of the spin parameter $a_{\rm bh}$).
Also, the accretion rate is limited to a multiple 
of the  Eddington mass accretion rate $\dot{M}_{\rm Edd}$
by a free parameter $A_{\rm Edd}$, which we set to $A_{\rm Edd}=1$ for heavy seeds, and to $A_{\rm Edd}\approx2.2$ for light seeds (so as to allow for mildly super-Eddington accretion~\cite{supereddington}). 
The bolometric quasar luminosity is then simply determined via $ L_{\rm bol}=\dot{M}_{\rm bh,\,QSO} c^2$.

\section{Details of the statistical method}
\label{app:statistics}

\begin{figure}
\begin{center}
	\includegraphics[width=.9\textwidth]{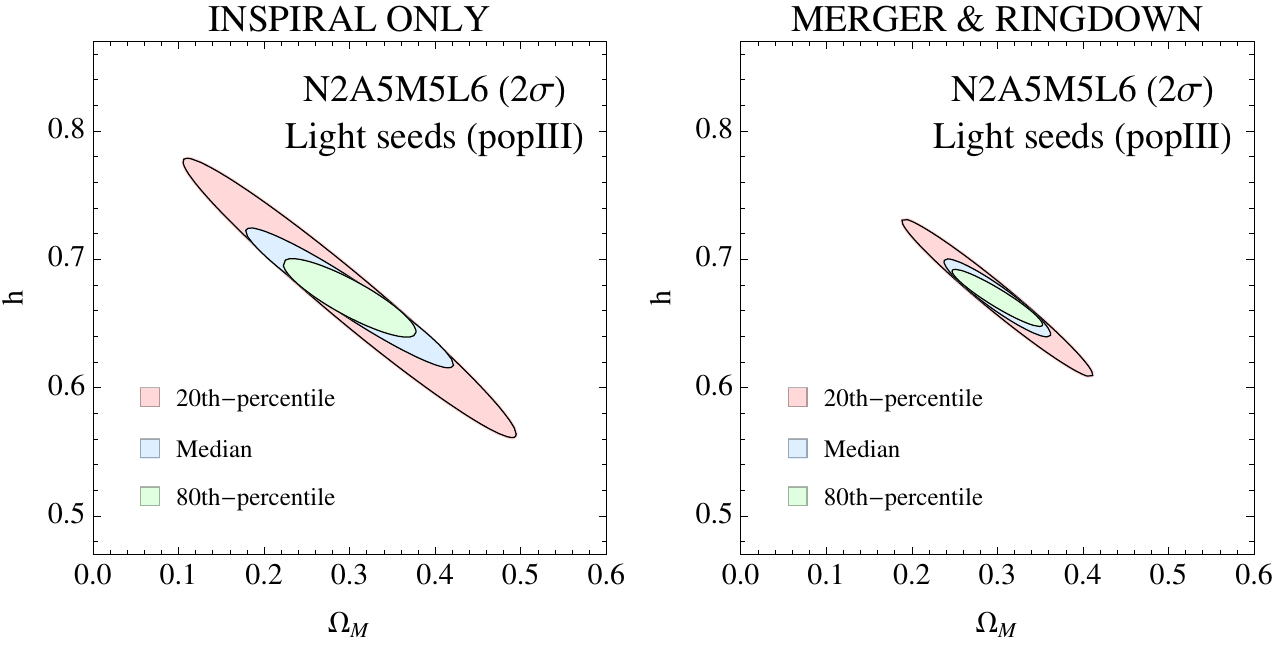}
\end{center}
\caption{2$\sigma$ likelihood contours derived from the Fisher matrix which gives an FoM closest to the median of the FoM PDF (used in all previous results), and to the Fisher matrices which give FoMs closest to its 20th and 80th percentile. The plot is for $\Lambda$CDM, the configuration N2A5M5L6, and the popIII model.}
\label{fig:LCDM_percentile}
\end{figure}

As discussed in section \ref{sec:statistic}, when drawing likelihood contour plots for the cosmological parameters, 
we have always shown the results obtained from the Fisher matrix that gives the closest FoM 
to the median of the FoM PDF. The models of MBHB formation discussed in 
section~\ref{sec:BHbinaries} provide source catalogues with an intrinsic, unavoidable scatter 
due to the stochastic nature of galaxy and massive BH formation and evolution. 
We have therefore made a choice for the Fisher matrix that in our opinion could best 
represent the performance of a given eLISA design. 
To have an idea of the possible scatter of the constraints on the cosmological parameters, 
in Fig.~\ref{fig:LCDM_percentile} we plot the likelihood contours 
derived from the Fisher matrix that gives the closest FoM to the median of the FoM PDF, and from the Fisher 
matrices that give the closest FoM to the 20th and 80th percentiles. 
The plots are for the configuration N2A5M5L6 and for the PopIII model in both the conservative
 (inspiral only) and optimistic (inspiral, merger and ringdown) scenarios. 
 We have decided to show this figure for the $\Lambda$CDM case, 
 since it is the most representative of the present situation in cosmology.
As  can be seen from Fig.~\ref{fig:LCDM_percentile}, the constraints on the parameter space might be either
slightly better or slightly worse than those obtained with the median Fisher matrix.
Note that the degeneracy between $\Omega_M$ and $h$ changes when going 
from worse to better measurements, implying that catalogues with higher FoM 
(which better constrain the parameter space) are more efficient at measuring 
some of the parameters (e.g.~$h$ for $\Lambda$CDM) than others (e.g.~$\Omega_M$).

Although we have simulated 118 catalogues for each MBHB formation scenario in order to improve our statistics,
the scatter of the properties of the MBHB population across the catalogues is still quite high. 
This can be understood by looking at the probability distribution of the FoM. In Fig.~\ref{fig:PDF} we plot the PDFs 
 of the FoM of the 118 catalogues, for each MBHB formation scenario, 
 two eLISA configurations and two cosmological models ($\Lambda$CDM and DDE), 
 in both the optimistic and conservative scenarios. The FoM is defined as in Eq.~\eqref{FoM}.
Naturally, the distributions are more peaked (and their peaks are centred on higher values of the FoM) 
for the cosmological models characterised by fewer parameters.
The Q3d model gives systematically the worst FoM for any eLISA configuration and any cosmological model, because
 fewer BH binaries are formed in this model than in the other two (see Fig.~\ref{fig:ss}). 
 The 90th and 10th percentiles of the FoM PDF 
 capture essentially all the scatter in the FoM, and we therefore quote these values 
 when presenting the results in section~\ref{sec:FoM}. 
\begin{figure}
\begin{center}
	\includegraphics[width=\textwidth]{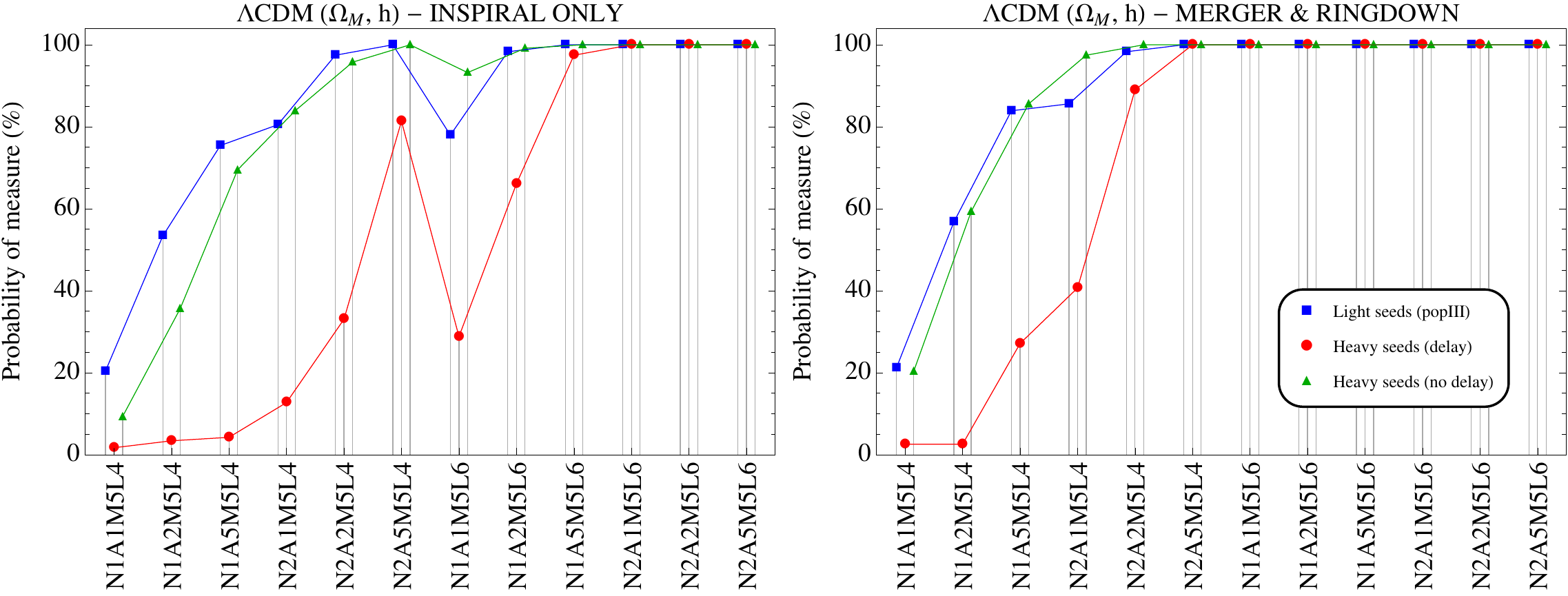}
\end{center}
\caption{Percentage of catalogues giving finite errors (i.e.~the probability of constraining at some finite level the cosmological parameters) in the $\Lambda$CDM model for the different eLISA configurations.}
\label{fig:PoM}
\end{figure} 
\begin{figure}
\begin{center}
	\includegraphics[width=.85\textwidth]{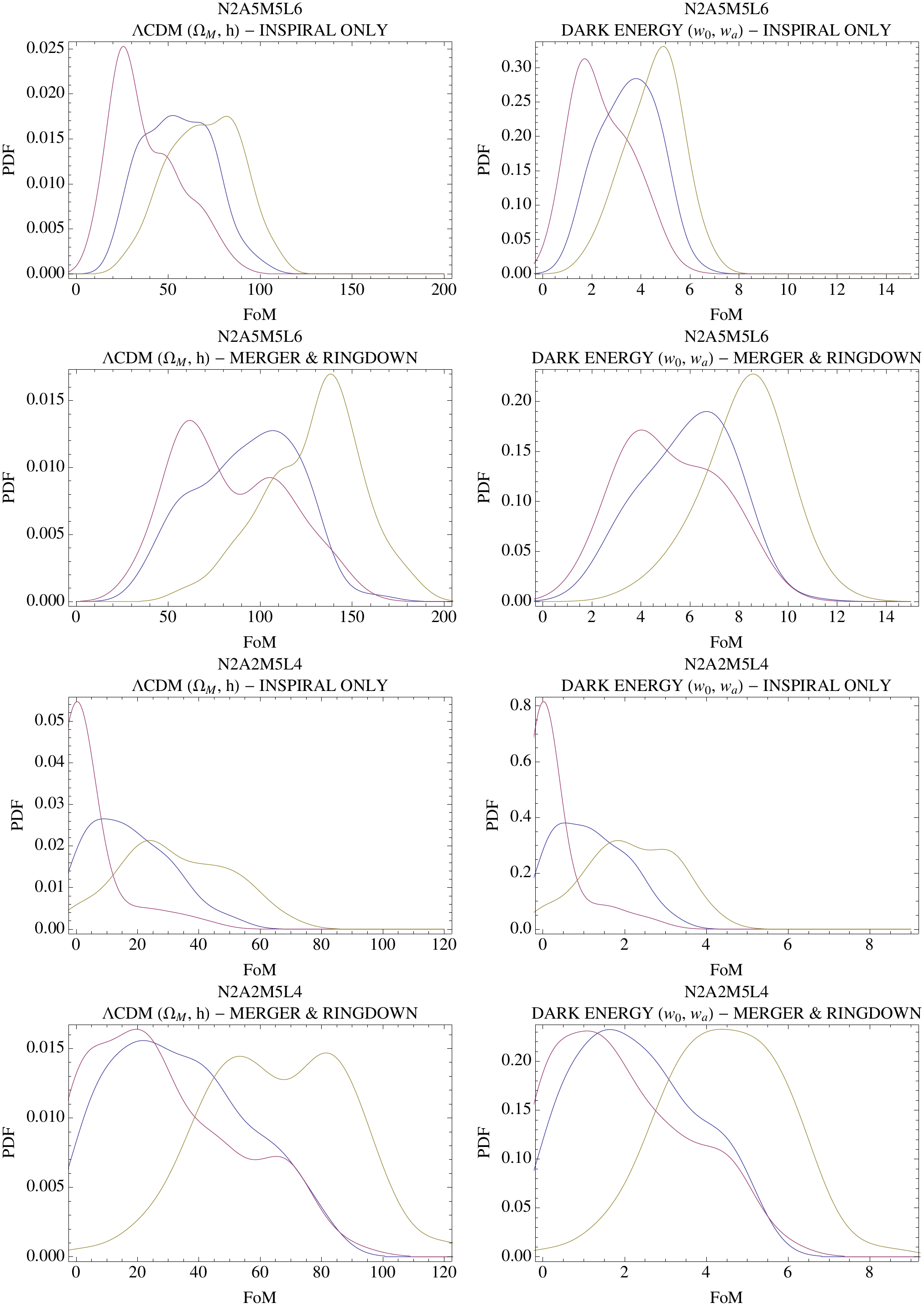}
\end{center}
\caption{Probability distribution functions for the FoM of N2A5M5L6 and N2A2M5L4. The three curves show the three astrophysical models of MBHB formation and evolution: popIII in blue, Q3d in red and Q3nod in yellow. We show the results for the $\Lambda$CDM and DDE cosmological models (two-parameter models) in both the conservative (inspiral only) and optimistic (inspiral plus merger and ringdown) scenarios.}
\label{fig:PDF}
\end{figure}
  
As discussed above, when we define one single FoM per configuration, 
cosmological model and MBHB formation scenario, we choose the median of the FoM PDF (shown in Fig.~\ref{fig:PDF}). 
This is useful to compare different eLISA configurations. However, when we need to plot likelihood contours, 
we select a single covariance matrix for each eLISA configuration/MBHB formation scenario/cosmological model, namely 
the one that gives the closest FoM to the median of the FoM PDF. This is done first 
in the five-parameter cosmological model. When we reduce 
to the three-parameter model of $\Lambda$CDM with curvature, 
we re-calculate the  FoM PDF by using all the Fisher matrices $F_{ij}^C$, in which we
impose an exact prior on $w_0$ and $w_a$ by deleting the corresponding
rows and columns, and then find again the median. The same procedure is used
 for each two-parameter model analysed, i.e.~we always re-calculate the FoM PDF, 
 starting from the complete set of the Fisher matrices with the relevant rows and columns removed.

To provide instead an estimate of the 1$\sigma$ standard errors on the cosmological parameters, 
we consider the median of the errors resulting from all the 118 catalogues pertaining to a given
MBHB formation model and cosmological model (and for a given eLISA design), as explained in section~\ref{sec:statistic}.
For some of the eLISA configurations (notably the L4 ones), the number of standard sirens is not 
enough to constrain the parameters of the cosmological model.
These cases are characterised by ill-defined Fisher matrices (with negative eigenvalues), 
and usually appear when the number of data points is lower than the number of cosmological parameters.
In these situations, the covariance matrix cannot be defined in a consistent manner,
 and the standard errors cannot be derived in the usual way.
For this reason, as mentioned in section~\ref{sec:statistic}, when the low number of standard sirens does not allow one to 
constrain the cosmological parameters, we simply assume infinite errors on them, and we do not take  
into account the corresponding catalogues when calculating the median FoM.
The resulting representative errors (cf.~Table~\ref{tab:standard_errors} 
and Table~\ref{tab:standard_errors_opt}) must thus always be considered in combination 
with the probability of measurement $P$, i.e.~the percentage of catalogues that actually provide non-infinite errors.
A comparison between the probability of measurement $P$ of all the different eLISA configurations is shown 
in Fig.~\ref{fig:PoM}, for both the conservative and optimistic scenarios.
Looking at Fig.~\ref{fig:PoM}, on the one hand it is clear that for six-link
 configurations, especially in the optimistic scenario, there are few catalogues 
 for which the cosmological parameters cannot be constrained.
On the other hand, in the worst four-link configurations the majority of catalogues cannot provide any constraints
 on the cosmological parameters, while the situation improves for the best designs, which have 
  comparable performance to six-link configurations in the optimistic scenario.
As expected, the
Q3d model always gives systematically lower probability $P$ than the other two models, 
due to the lower number of detectable standard sirens (cf.~Fig.~\ref{fig:ss}).

\begin{table}
\begin{center}
\begin{tabular}{|c|cc|cc|cc|}
	\hline
	& \multicolumn{2}{|c|}{20th perc.} & \multicolumn{2}{|c|}{50th perc.} & \multicolumn{2}{|c|}{80th perc.} \\
	\hline
	& $\Delta \Omega_M$ & $\Delta h$ & $\Delta \Omega_M$ & $\Delta h$ & $\Delta \Omega_M$ & $\Delta h$ \\
	\hline
	\multirow{2}{*}{INSPIRAL} & 0.0784 & 0.0438 & 0.0489 & 0.0220 & 0.0313 & 0.0123 \\
	\multirow{2}{*}{ONLY} & 0.152 & 0.0900 & 0.0971 & 0.0478 & 0.0478 & 0.0201 \\
	& 0.0606 & 0.0280 & 0.0339 & 0.0128 & 0.0284 & 0.00968 \\
 	\hline
 	\multirow{2}{*}{MERGER \&} & 0.0450 & 0.0247 & 0.0251 & 0.0122 & 0.0213 & 0.00896 \\
 	\multirow{2}{*}{RINGDOWN} & 0.0535 & 0.0337 & 0.0338 & 0.0186 & 0.0184 & 0.00829 \\
 	& 0.0250 & 0.0127 & 0.0168 & 0.00737 & 0.0162 & 0.00688 \\
 	\hline
\end{tabular}
\end{center}
\caption{Scatter of the errors of the parameters of the $\Lambda$CDM model for both conservative (inspiral only) and optimistic (merger \& ringdown) scenarios with N2A5M5L6 (cf.~Fig.~\ref{fig:LCDM_percentile}). The errors in the middle column (50\% percentile, given inverting the covariance matrix which provides an FoM closest to the median of the FoM PDF) are only slightly different from the median errors, those of the fourth column of Table~\ref{tab:standard_errors}.}
\label{tab:errors_scatter}
\end{table}

Finally, in order to compare the median of the errors and the errors obtained from the covariance matrix whose 
FoM is closest to the median of the FoM distribution (i.e.~the one used to plot likelihood contours), 
in Table~\ref{tab:errors_scatter} we show the errors obtained from 
the covariance matrices used in Fig.~\ref{fig:LCDM_percentile}.
Comparing the 50th percentile (central column) values in Table~\ref{tab:errors_scatter} with the corresponding ones 
in Table~\ref{tab:standard_errors}, one can see that the former well represent the latter, 
at least for the $\Lambda$CDM cosmological model.
Furthermore, Table~\ref{tab:errors_scatter} gives an 
idea of the possible scatter that we can obtain within the same cosmological model and MBHB formation scenario.
In fact, the errors obtained from the nearest covariance matrix to the 20th and 80th percentile of the 
FoM distribution are presented together with the 50th percentile (i.e.~the median).
We see from Table~\ref{tab:errors_scatter} that the errors vary at most by a factor of 5 in both conservative and optimistic scenarios.

\bibliographystyle{unsrt}
\bibliography{COSMO_paper}

\newpage

\begin{table}
\begin{center}
\centerline{\begin{tabular}{|c|c|c|c|ccc|ccc|c|}
	\hline
	 & SNR & $\Delta\Omega<$ & \multirow{2}{*}{LSST} & \multicolumn{3}{c|}{SKA (radio only)} & \multicolumn{3}{c|}{SKA + ELT} & \multirow{2}{*}{Total} \\
	 & $>$ 8 & $10\,{\rm deg}^2$ &  & Flare & Jet & Total & Spec & Photo & Total & \\
	\hline
	\multirow{3}{*}{N1A1M5L4} 
							  & 28.8 & 0.569 & 0.0259 & 0.534 & 0.431 & 0.534 & 0.388 & 0.147 & 0.534 & 0.534 \\
							  & 11.9 & 0.0517 & 0.0172 & 0.0517 & 0.0259 & 0.0517 & 0.0517 & 0 & 0.0517 & 0.0517 \\
							  & 90.9 & 0.431 & 0 & 0.319 & 0.00862 & 0.319 & 0.259 & 0.0603 & 0.319 & 0.319 \\
	\hline
	\multirow{3}{*}{N1A2M5L4} 
							  & 69.0 & 1.54 & 0.0431 & 1.44 & 1.23 & 1.44 & 0.836 & 0.526 & 1.36 & 1.36 \\
							  & 21.8 & 0.0517 & 0.0172 & 0.0517 & 0.0172 & 0.0517 & 0.0517 & 0 & 0.0517 & 0.0517 \\
							  & 178 & 1.39 & 0.00862 & 0.914 & 0.0259 & 0.914 & 0.655 & 0.259 & 0.914 & 0.914 \\
	\hline
	\multirow{3}{*}{N1A5M5L4} 
							  & 150 & 2.96 & 0.0948 & 2.67 & 2.35 & 2.72 & 1.42 & 1.14 & 2.56 & 2.56 \\
							  & 33.7 & 0.181 & 0.0603 & 0.181 & 0.129 & 0.181 & 0.172 & 0.00862 & 0.181 & 0.181 \\
							  & 336 & 3.61 & 0.0172 & 2.01 & 0.103 & 2.02 & 1.40 & 0.586 & 1.98 & 1.98 \\
	\hline
	\multirow{3}{*}{N2A1M5L4} 
							  & 93.7 & 2.92 & 0.112 & 2.60 & 2.09 & 2.62 & 1.68 & 0.871 & 2.55 & 2.55 \\
							  & 37.4 & 0.328 & 0.112 & 0.328 & 0.181 & 0.328 & 0.319 & 0.00862 & 0.328 & 0.328 \\
							  & 415 & 4.39 & 0.0517 & 2.63 & 0.164 & 2.64 & 1.83 & 0.733 & 2.56 & 2.56 \\
	\hline
	\multirow{3}{*}{N2A2M5L4} 
							  & 229 & 5.87 & 0.224 & 5.16 & 4.63 & 5.39 & 2.65 & 2.35 & 5.00 & 5.00 \\
							  & 39.4 & 0.836 & 0.353 & 0.836 & 0.543 & 0.836 & 0.793 & 0.0345 & 0.828 & 0.828 \\
							  & 497 & 8.97 & 0.155 & 5.25 & 0.474 & 5.29 & 3.28 & 1.82 & 5.10 & 5.10 \\
	\hline
	\multirow{3}{*}{N2A5M5L4} 
							  & 520 & 12.9 & 0.371 & 11.1 & 11.3 & 12.3 & 4.61 & 5.83 & 10.4 & 10.4 \\
							  & 40.6 & 2.77 & 1.03 & 2.75 & 1.91 & 2.77 & 2.22 & 0.534 & 2.75 & 2.75 \\
							  & 557 & 18.7 & 0.534 & 11.1 & 1.39 & 11.2 & 5.95 & 4.15 & 10.1 & 10.1 \\
	\hline
	\multirow{3}{*}{N1A1M5L6} 
							  & 48.2 & 2.90 & 0.155 & 2.72 & 2.26 & 2.73 & 1.84 & 0.845 & 2.68 & 2.68 \\
							  & 17.1 & 0.741 & 0.267 & 0.741 & 0.543 & 0.741 & 0.612 & 0.129 & 0.741 & 0.741 \\
							  & 133.8 & 5.27 & 0.0776 & 4.42 & 0.241 & 4.42 & 2.10 & 1.71 & 3.81 & 3.81 \\
	\hline
	\multirow{3}{*}{N1A2M5L6} 
							  & 116 & 7.38 & 0.216 & 6.84 & 6.29 & 7.03 & 3.05 & 3.24 & 6.29 & 6.29 \\
							  & 27.7 & 1.98 & 0.569 & 1.97 & 1.47 & 1.98 & 1.36 & 0.560 & 1.92 & 1.92 \\
							  & 245 & 14.0 & 0.0862 & 9.67 & 0.612 & 9.68 & 3.36 & 4.10 & 7.47 & 7.47 \\
	\hline
	\multirow{3}{*}{N1A5M5L6} 
							  & 244 & 17.4 & 0.379 & 14.7 & 15.8 & 16.8 & 5.55 & 8.69 & 14.2 & 14.2 \\
							  & 37.7 & 4.91 & 1.09 & 4.84 & 3.79 & 4.91 & 2.66 & 1.94 & 4.59 & 4.59 \\
							  & 417 & 39.2 & 0.224 & 21.9 & 2.45 & 22.0 & 5.53 & 8.96 & 14.5 & 14.5 \\
	\hline
	\multirow{3}{*}{N2A1M5L6} 
							  & 152 & 13.6 & 0.543 & 12.1 & 11.8 & 13.0 & 5.67 & 6.22 & 11.9 & 11.9 \\
							  & 39.9 & 7.18 & 1.51 & 7.13 & 5.34 & 7.18 & 4.00 & 2.71 & 6.71 & 6.71 \\
							  & 499 & 49.2 & 0.431 & 27.8 & 3.55 & 27.9 & 6.92 & 10.9 & 17.8 & 17.8 \\
	\hline
	\multirow{3}{*}{N2A2M5L6} 
							  & 359 & 30.8 & 0.724 & 24.1 & 28.7 & 30.1 & 8.34 & 15.3 & 23.7 & 23.7 \\
							  & 40.8 & 11.9 & 1.97 & 11.7 & 8.54 & 11.9 & 5.59 & 5.31 & 10.9 & 10.9 \\
							  & 561 & 85.7 & 0.603 & 39.9 & 5.43 & 40.1 & 8.82 & 15.0 & 23.8 & 23.8 \\
	\hline
	\multirow{3}{*}{N2A5M5L6} 
							  & 682 & 71.5 & 0.940 & 46.9 & 69.1 & 70.7 & 12.1 & 30.0 & 42.1 & 42.1 \\
							  & 41.0 & 16.8 & 2.78 & 16.4 & 12.4 & 16.8 & 7.68 & 7.42 & 15.1 & 15.1 \\
							  & 595 & 149. & 1.03 & 60.8 & 9.16 & 61.2 & 11.6 & 20.8 & 32.4 & 32.4 \\
	\hline
\end{tabular}}
\end{center}
\captionsetup{width=1.12\textwidth}
\caption{In the conservative scenario (inspiral only), average values (5 years) of counterpart detections for all eLISA configurations and all three MBHB formation models: popIII, Q3d and Q3nod (respectively from top to bottom in each cell). From left to right the table shows: the eLISA detections (inspiral only), the eLISA detections with sky location error below 10 deg$^2$ (inspiral only), the optical counterparts observed with LSST, the radio counterparts observed with SKA (respectively from the flare, jet and both emissions combined), the optical observations with ELT of SKA counterparts hosts (with redshift measured spectroscopically and photometrically) and the total useful standard sirens (LSST + ELT).}
\label{tab:ss}
\end{table}

\newpage

\begin{table}
\begin{center}
\centerline{\begin{tabular}{|c|c|c|c|ccc|ccc|c|}
	\hline
	 & SNR & $\Delta\Omega<$ & \multirow{2}{*}{LSST} & \multicolumn{3}{c|}{SKA (radio only)} & \multicolumn{3}{c|}{SKA + ELT} & \multirow{2}{*}{Total} \\
	 & $>$ 8 & $10\,{\rm deg}^2$ &  & Flare & Jet & Total & Spec & Photo & Total & \\
	\hline
	\multirow{3}{*}{N1A1M5L4} 
							  & 39.1 & 0.759 & 0.0431 & 0.664 & 0.448 & 0.664 & 0.517 & 0.147 & 0.664 & 0.664 \\
							  & 37.8 & 0.0776 & 0.0259 & 0.0776 & 0.0345 & 0.0776 & 0.0776 & 0 & 0.0776 & 0.0776 \\
							  & 370.6 & 0.681 & 0 & 0.466 & 0.00862 & 0.466 & 0.405 & 0.0603 & 0.466 & 0.466 \\
	\hline
	\multirow{3}{*}{N1A2M5L4} 
							  & 75.7 & 1.85 & 0.0862 & 1.63 & 1.28 & 1.63 & 1.03 & 0.526 & 1.55 & 1.55 \\
							  & 39.9 & 0.129 & 0.0776 & 0.129 & 0.0776 & 0.129 & 0.129 & 0 & 0.129 & 0.129 \\
							  & 488 & 2.23 & 0.0517 & 1.34 & 0.0948 & 1.34 & 1.08 & 0.259 & 1.34 & 1.34 \\
	\hline
	\multirow{3}{*}{N1A5M5L4} 
							  & 152 & 3.47 & 0.233 & 2.98 & 2.56 & 3.07 & 1.77 & 1.15 & 2.91 & 2.91 \\
							  & 40.7 & 0.647 & 0.267 & 0.647 & 0.414 & 0.647 & 0.586 & 0.0603 & 0.647 & 0.647 \\
							  & 566 & 5.33 & 0.181 & 3.11 & 0.362 & 3.14 & 2.37 & 0.716 & 3.09 & 3.09 \\
	\hline
	\multirow{3}{*}{N2A1M5L4} 
							  & 95.8 & 3.52 & 0.302 & 3.06 & 2.37 & 3.11 & 2.16 & 0.879 & 3.04 & 3.04 \\
							  & 40.7 & 1.16 & 0.534 & 1.16 & 0.767 & 1.16 & 1.05 & 0.112 & 1.16 & 1.16 \\
							  & 573 & 6.75 & 0.328 & 4.29 & 0.690 & 4.34 & 3.33 & 0.905 & 4.23 & 4.23 \\
	\hline
	\multirow{3}{*}{N2A2M5L4} 
							  & 230 & 7.03 & 0.638 & 6.10 & 5.46 & 6.47 & 3.69 & 2.40 & 6.09 & 6.09 \\
							  & 40.9 & 3.45 & 1.54 & 3.44 & 2.35 & 3.45 & 2.91 & 0.500 & 3.41 & 3.41 \\
							  & 595 & 13.7 & 0.690 & 8.66 & 1.71 & 8.79 & 5.87 & 2.54 & 8.41 & 8.41 \\
	\hline
	\multirow{3}{*}{N2A5M5L4} 
							  & 520 & 15.7 & 1.00 & 13.0 & 13.6 & 15.0 & 6.98 & 6.11 & 13.1 & 13.1 \\
							  & 41.0 & 14.7 & 3.47 & 14.6 & 10.3 & 14.7 & 8.82 & 5.48 & 14.3 & 14.3 \\
							  & 608 & 43.8 & 1.25 & 27.0 & 5.57 & 27.3 & 11.2 & 10.6 & 21.9 & 21.9 \\
	\hline
	\multirow{3}{*}{N1A1M5L6} 
							  & 57.7 & 8.15 & 0.862 & 6.71 & 6.36 & 7.52 & 5.59 & 1.88 & 7.47 & 7.47 \\
							  & 39.9 & 10.0 & 2.09 & 10.0 & 7.66 & 10.0 & 5.34 & 4.36 & 9.70 & 9.70 \\
							  & 461 & 32.2 & 0.724 & 24.7 & 4.07 & 24.8 & 7.49 & 9.30 & 16.8 & 16.8 \\
	\hline
	\multirow{3}{*}{N1A2M5L6} 
							  & 121 & 13.9 & 1.04 & 11.3 & 11.8 & 13.2 & 7.56 & 4.92 & 12.5 & 12.5 \\
							  & 40.8 & 15.1 & 2.71 & 15.0 & 11.1 & 15.1 & 7.33 & 6.85 & 14.2 & 14.2 \\
							  & 555 & 72.9 & 0.966 & 43.3 & 6.68 & 43.5 & 10.1 & 14.9 & 25.0 & 25.0 \\
	\hline
	\multirow{3}{*}{N1A5M5L6} 
							  & 246 & 24.3 & 1.17 & 18.8 & 22.1 & 23.6 & 10.2 & 10.8 & 21.0 & 21.0 \\
							  & 41.0 & 21.0 & 3.71 & 20.8 & 15.3 & 21.0 & 9.70 & 9.70 & 19.4 & 19.4 \\
							  & 599 & 142. & 1.28 & 64.5 & 10.2 & 65.2 & 12.7 & 20.4 & 33.1 & 33.1 \\
	\hline
	\multirow{3}{*}{N2A1M5L6} 
							  & 153 & 18.5 & 1.18 & 15.1 & 16.2 & 17.7 & 9.29 & 7.28 & 16.6 & 16.6 \\
							  & 41.0 & 23.4 & 4.16 & 23.2 & 17.1 & 23.4 & 11.1 & 10.6 & 21.7 & 21.7 \\
							  & 602 & 151. & 1.54 & 68.5 & 11.1 & 69.1 & 13.7 & 20.8 & 34.5 & 34.5 \\
	\hline
	\multirow{3}{*}{N2A2M5L6} 
							  & 360 & 35.4 & 1.24 & 26.8 & 32.9 & 34.7 & 11.7 & 16.6 & 28.2 & 28.2 \\
							  & 41.1 & 28.9 & 4.37 & 28.5 & 20.4 & 28.9 & 12.6 & 13.8 & 26.4 & 26.4 \\
							  & 610 & 214. & 1.79 & 84.5 & 13.5 & 85.3 & 15.3 & 25.1 & 40.5 & 40.5 \\
	\hline
	\multirow{3}{*}{N2A5M5L6} 
							  & 683 & 75.2 & 1.28 & 49.0 & 72.7 & 74.5 & 14.6 & 31.2 & 45.8 & 45.8 \\
							  & 41.1 & 35.3 & 4.50 & 34.7 & 25.1 & 35.3 & 13.6 & 17.6 & 31.2 & 31.2 \\
							  & 611 & 385. & 1.88 & 127. & 18.5 & 128. & 16.7 & 33.3 & 50.0 & 50.0 \\
	\hline
\end{tabular}}
\end{center}
\captionsetup{width=1.12\textwidth}
\caption{The same as Table \ref{tab:ss}, but in the optimistic scenario (with merger and ringdown and de-lensing of a factor of two).}
\label{tab:ss2}
\end{table}

\end{document}